%% file: main.tex
\documentclass[twocolumn]{aastex631}

\usepackage{graphicx}
\usepackage[caption=false]{subfig}
\DeclareCaptionLabelFormat{continued}{#1 ̃#2 (cont.)}
\usepackage[figure,figure*]{hypcap}

\newcommand{\teff}{$T_{\mathrm{eff}}$}
\newcommand{\logg}{$\log(g)$}

\received{April 29, 2021}
\revised{July 14, 2021}
\accepted{July 16, 2021}

\submitjournal{\apj}

\shorttitle{ZTF Variability in Dwarf Carbon Stars}
\shortauthors{Roulston et al.}

\turnoffediting

\begin{document}

\title{Unexpected Short-Period Variability in Dwarf Carbon Stars from the Zwicky Transient Facility}

\correspondingauthor{Benjamin Roulston}
\email{broulston@cfa.harvard.edu}

\author[0000-0002-9453-7735]{Benjamin R. Roulston}
\altaffiliation{SAO Predoctoral Fellow}
\affiliation{Center for Astrophysics $\vert$ Harvard \& Smithsonian, 60 Garden Street, Cambridge, MA 02138, USA}
\affiliation{Department of Astronomy, Boston University, 725 Commonwealth Avenue, Boston, MA 02215, USA}

\author[0000-0002-8179-9445]{Paul J. Green}
\affiliation{Center for Astrophysics $\vert$ Harvard \& Smithsonian, 60 Garden Street, Cambridge, MA 02138, USA}

\author[0000-0002-2998-7940]{Silvia Toonen}
\affiliation{Anton Pannekoek Institute, University of Amsterdam, Science Park 904, 1098 XH Amsterdam, Netherlands}

\author[0000-0001-5941-2286]{J. J. Hermes}
\affiliation{Department of Astronomy, Boston University, 725 Commonwealth Avenue, Boston, MA 02215, USA}



\begin{abstract}
Dwarf carbon (dC) stars, main sequence stars showing carbon molecular bands, are enriched by mass transfer from a previous asymptotic-giant-branch (AGB) companion, which has since evolved to a white dwarf. While previous studies have found radial-velocity variations for large samples of dCs, there are still relatively few dC orbital periods in the literature and no dC eclipsing binaries have yet been found. Here, we analyze photometric light curves from DR5 of the Zwicky Transient Facility for a sample of 944 dC stars. From these light curves, we identify 34 periodically variable dC stars. Remarkably, of the periodic dCs, 82\% have periods less than two days. We also provide spectroscopic follow-up for four of these periodic systems, measuring radial velocity variations in three of them. Short-period dCs are almost certainly post-common-envelope binary systems, since the periodicity is most likely related to the orbital period, with tidally locked rotation and photometric modulation on the dC either from spots or from ellipsoidal variations. We discuss evolutionary scenarios that these binaries may have taken to accrete sufficient C-rich material while avoiding truncation of the thermally pulsing AGB phase needed to provide such material in the first place. We compare these dCs to common-envelope models to show that dC stars probably cannot accrete enough C-rich material during the common-envelope phase, suggesting another mechanism like wind-Roche lobe overflow is necessary. The periodic dCs in this paper represent a prime sample for spectroscopic follow-up and for comparison to future models of wind-Roche lobe overflow mass transfer.
\end{abstract}



\section{Introduction}\label{sec:intro}

Carbon (C) stars are those that show molecular absorption bands of C, such as C$_2$, CN and CH, in their optical spectra \citep{Secchi1869}. Intrinsic C stars have experienced C enrichment via dredge-up of fusion products from their cores. During the thermally pulsing (TP) phase of the asymptotic giant branch (AGB), shell He flashes cause strong convection in the shell regions bringing C into the atmosphere --- the third dredge-up. If the C/O ratio increases above unity, a giant C star is formed, since C preferentially binds with oxygen to form CO, leaving excess C to form the C molecules of C$_2$, CN and CH. Thus, it was long thought that all C stars were giants on the TP-AGB.

This made it quite surprising when \citet{Dahn1977} found the first \textit{main-sequence} C star, G77-61. This dwarf carbon (dC) star cannot have yet experienced fusion to create C enhancement, or the third dredge-up, as it is a main-sequence star. \citet{Dahn1977} put forth a few explanations of this C enhancement on the main-sequence, with the favored being that G77-61 was in a binary system and had experienced C enhanced mass transfer from a previous AGB companion. This former AGB companion would since have become a white dwarf (WD) and cooled until no longer detectable alongside the dC. Indeed this seemed to have been confirmed when \cite{Dearborn1986} found G77-61 to be a radial velocity (RV) binary with an orbital period of 245.5\,d.

Today, G77-61 is no longer the only known dC, with close to 1000 known in the literature. The majority of these dCs come from the \citet{Green2013} and \citet{Si2014} samples which were found from all-sky spectroscopic surveys. This has included almost a dozen ``smoking gun'' systems in which the WD companion is sufficiently hot to be visible in the optical spectra as a spectroscopic composite \citep{Heber1993, Liebert1994, Green2013, Si2014}. These samples have shown that the dC stars are actually the most common type of C star in the Galaxy. 

Their carbon-enriched atmospheres make dCs the most likely progenitors of the carbon-enhanced metal-poor, CH, and possibly the Ba II stars, all showing carbon and \textit{s}-process enhancements \citep{Jorissen2016, DeMarco2017}. These stars, being more luminous than dCs, have been studied via RV campaigns, which have shown increased binarity compared to normal O-rich stars, indicating they have likely also experienced mass transfer from an unseen companion \citep{Sperauskas2016}. Barium dwarfs and CH subgiants show periods from RV analysis of 1--20\,years \citep{Escorza2019}.

Blue straggler stars are another class similar to dCs in that they may have experienced mass transfer from a previous AGB companion. As discussed by \citet{Gosnell2019}, blue straggler stars in a cluster color-magnitude diagram are more luminous and bluer than the main sequence turnoff. While some are likely formed in mergers or collisions, most blue straggler stars are thought, like dC stars, to be the result of mass transfer from a giant to a main sequence dwarf.  Most blue straggler stars are found in wide binaries with periods of order 1000 days, consistent with expectations of mass transfer from an AGB star onto a main-sequence companion \citep{Chen2008, Gosnell2014},  which leaves a CO-core WD remnant \citep{Paczynski1971}. Those blue straggler stars that form after mass transfer from an red giant branch star, yields a blue straggler in a shorter binary period \citep[of order 100 days;][]{Chen2008} leaving a He-core WD companion.  The salient point relevant to dC stars is that significant mass is typically gained in these encounters.

While dC stars are now known to be numerous, details of their properties remains sparse. This is especially true of their orbital properties. Currently, there are only six orbital periods for dCs in the literature. The first is of the dC prototype G77-61, found to be a single-line spectroscopic binary with an orbital period of 245.5\,d \citep{Dearborn1986}. The central source of the Necklace Nebula was found to be a binary with a dC, which has a photometric period of 1.16\,d \citep{Corradi2011, Miszalski2013}. The three longest period dCs in the literature are those from \citet{Harris2018} who found astrometric binaries with periods of 1.23\,yr, 3.21\,yr, and 11.35\,yr. \citet{Margon2018} found a dC with a photometric period of 2.92\,d and confirmed this as the orbital period with spectroscopic follow-up. 

\added{In their recent work, \citet{Whitehouse2021} conducted an RV survey of seven dCs with H$\alpha$ emission, finding short orbital periods for all of them (six new periods). In addition, they found photometric periods with similar lengths as the orbital periods in the range of 0.2--5.2\,d. Their light curve modeling suggests that the source of variability in their dCs is from stellar rotation and spots. As with the new photometrically periodic dCs in this paper, these dCs must have experienced a common-envelope phase with the former AGB companion.}

There have also been large sample few-epoch spectroscopy campaigns of dCs. \citet{Whitehouse2018} conducted a few-epoch survey of 28 dCs finding RV variability in 21 of them, implying a high binary fraction. \citet{Roulston2019} conducted a larger survey of 240 dCs using few-epoch spectroscopy from the Sloan Digital Sky Survey \citep[SDSS;][]{SDSS_4}. They found that dCs are consistent with 100\% binarity, with separations of order 1\,au and periods of order 1\,yr. Both the \citet{Whitehouse2018} and \citet{Roulston2019} surveys lacked enough spectral epochs to fit individual orbits and instead relied on statistical analysis to describe the dC population as a whole.

 \citet{Kool1995} modeled the space density of dCs, and they predicted the dC period to be bimodal with peaks near $10^2$--$10^3$\,d and $10^3$--$10^5$\,d, consistent with the known periods listed above. \citet{Kool1995} also found that the production of dCs is strongly dependent on metallicity, finding no dCs should be formed in systems with initial metallicity greater than half of solar (i.e. [Fe/H]$ > -0.3$). This is in agreement with metallicity measurements of G77-61, where \citet{Plez2005} found [Fe/H] = $-4.0$, making G77-61 one of the lowest metallicity stars known. This also has been supported by \citet{Farihi2018} who found 30--60\% of dCs to be halo objects, which are metal poor.

Until this year, the known dC periods spanned from $\sim1$\,d to $\sim4100$\,d, likely indicating different formation pathways. The longest dC periods that are of order 10s of years likely experienced only standard Roche-lobe overflow (RLOF) or wind-RLOF (WRLOF). These periods are consistent with other types of post-mass-transfer systems, such as the blue straggler stars. 

The dCs with periods $\la1$\,d would have likely experienced common-envelope (CE) evolution \citep{Paczynski1976, Ivanova2013}, since the TP-AGB envelope expands to several 100s of solar radii. Of interest is how CE evolution affects dC formation. \deleted{From the known dC periods, it is very likely that dCs with periods near a few days or less will experience some form of CE during the AGB phase of the evolved companion. } Once the CE phase has started, the plunge-in of the lower-mass companion (in our case, the future dC) could truncate the evolution of the AGB by ejection of its envelope. If this happens before the TP-AGB phase and the third dredge-up, it is likely that the C enhancement needed for dC formation will not occur. However, if the CE begins after the AGB companion has already become a C-giant, then it may be possible for the main-sequence companion to accrete enough C-rich material from the CE to become a dC (depending on the accretion efficiency). If the accretion efficiency is not high enough, however, the main-sequence companion will not accrete enough material from the CE alone, requiring some combination of CE evolution with efficient mass transfer before the CE phase that is sufficient to transform an O-rich main-sequence star into a C-rich dC. 

Significant accretion of mass and angular momentum from the AGB companion could result in significant spin-up and subsequent activity in some dCs \citep{Green2019}.  If there are dCs left in very tight orbits with the WD remnant, they may show tidally locked rotation periods (synchronous rotation), as well as tidal distortions causing ellipsoidal variations in photometric light curves. A search for periodicity in photometrically variable dCs could reveal some systems useful for constraining their evolution.

Another motivation to study variability in dCs is that no dC masses have yet been measured, because there are no known eclipsing dC systems. We can estimate dC masses from optical or infrared (IR) colors (see Section \ref{subsec:dc_bps_compare}), but these estimates have uncharacterized systematics, due to differences between normal O-rich stars and C stars in the optical and IR regions. 

This lack of eclipsing dC systems highlights the importance of photometric surveys to search for the first well-characterized eclipsing dC systems. These systems could, when combined with RV follow-up, provide the first reliable dC mass measurements, and help us understand more about the amount, and composition, of accreted mass needed to form a dC. 

\citet{Margon2018} conducted a search for periodic dCs using the Palomar Transient Factory \citep[PTF;][]{Law2009, Rau2009}, finding just one periodic dC. However, they clearly highlighted the potential for large photometric surveys to find periodic dCs, particularly dCs with short periods that should have experienced the strongest phases of CE mass transfer. 

 In this paper we report on a unique sample of close binary dCs --- implicating them as post-common envelope binaries (PCEBs) and likely pre-CVs --- discovered from their periodic photometric variability in the Zwicky Transient Facility. In Section~\ref{sec:sample} we describe the sample of dCs we selected to search and in Section~\ref{sec:lc_processing} we detail our process for cleaning and preparing the raw light curves. In Section~\ref{sec:periodic} we describe our process for finding which dCs have detected periodic signals. In Section~\ref{sec:rvs} we present spectroscopic follow-up for four of the periodic dCs in this paper. Finally, in Section~\ref{sec:BPS} we present comparisons of these short period dCs to binary population synthesis models to understand how a common-envelope phase relates to dC formation.
\section{Sample Selection}\label{sec:sample}

To search for variability in as many dCs as possible, we compiled a list of all dCs from the current literature. The largest contributor \added{(747 dCs, 79\%)} is the \citet{Green2013} sample of carbon stars from the SDSS. \added{We also selected a smaller number of dCs from \citet{Si2014}, who found 96 new dCs using a label propagation algorithm from SDSS DR8, and from  \citet{Li2018} who selected carbon stars from  the Large Sky Area Multi-Object Fiber Spectroscopic Telescope survey \citep[LAMOST;][]{LAMOST} using a machine learning approach.} Our resulting final sample consists of 944 dCs.

With our compiled sample, to ensure that any periodic candidate was indeed a dwarf carbon star, we used Gaia EDR3 parallaxes, proper motions \citep{GaiaEDR3} and distances \citep{GaiaEDR3_dist}. We required that each periodic C star  had M$_G > 5$\,mag from Gaia EDR3 based either on (1) significant parallax $\varpi / \varpi_{\rm{err}} > 5$ (27/34 periodic dCs) or (2) a significant proper motion ($\mu/\sigma_\mu > 5$) which sets an upper limit on the dC distance by limiting its transverse velocity to be less than an assumed Galactic escape velocity \citep{Smith2007} of about 600\,km/s  (7/34 periodic dCs).

\section{Light Curve Processing} \label{sec:lc_processing}

Using our list of dCs, we cross-matched our sample to the Zwicky Transient Facility \replaced{DR3}{DR5} \citep[ZTF;][]{ZTF_1,ZTF_2,ZTF_3}. We required a match to be within 2\arcsec\ of our target coordinates and each star having $\geq 10$ epochs in the available ZTF filters.

From the resulting matches detected within each filter, we grouped all sources within the match distance to ensure all epochs for each dC were included. The final sample of light curves resulted in \replaced{668 dCs with ZTF $g$ light curves, 712 dCs with ZTF $r$ light curves, and 406 dCs with ZTF $i$}{833 dCs with ZTF $g$ light curves, 867 dCs with ZTF $r$ light curves, and 554 dCs with ZTF $i$} light curves. For each light curve, we only used epochs for which the ZTF flag \texttt{catflags} $ == 0$ (no ZTF flags), ensuring every epoch is of a high quality. We summarize the light curve sample for each filter in Table~\ref{tab:lc_stats}.

\begin{deluxetable}{cDDDDD}
\tablecaption{Light Curve Statistics}
\tablewidth{1.0\textwidth}
\tablehead{\colhead{Filter} & \twocolhead{$N_{\rm{stars}}$} & \twocolhead{$<N_{\rm{epochs}}>$} & \twocolhead{$\sigma_{\rm{N_{epochs}}}$} & \twocolhead{$<\rm{mag}>$} & \twocolhead{$<\sigma_{\rm{mag}}>$}}
\decimals
\startdata
ZTF $g$ & 833 & 185 & 204 & 19.32 & 0.11 \\
ZTF $r$ & 867 & 269 & 237 & 18.07 & 0.05 \\
ZTF $i$ & 554 & 31  & 22  & 17.81 & 0.05 \\
\enddata
\tablecomments{Statistics of the light curves in the three ZTF filters. For each filter we report the number of stars, the mean number of epochs, the standard deviation of the number of epochs, the mean magnitude, and the mean magnitude error.} 
\end{deluxetable}
\label{tab:lc_stats}

We checked for any epochs which appear to be discrepant by performing an outlier removal on all the light curves. We first select from the raw light curve the brightest and faintest 5\% of epochs. Within these brightest and faintest 5\%, we calculate the median magnitude of each (i.e. the median of the 5\% brightest and 5\% faintest) and the mean error of that same brightest and faintest 5\%. We then removed any outliers that were 2$\sigma$ brighter than the median of the brightest 5\%, and removed those 2$\sigma$ fainter than the median of the faintest 5\%. If this selection dropped the number of epochs below 10, we removed that light curve from our analysis. This \replaced{treament}{treatment} rejects most artifacts without removing genuine astrophysical variability.

We checked the light curves for each dC, in each filter, to determine if each dC had detected variability by examining how the mean magnitude changed across the observed light curve time span. A small number of dCs which show no periodic variability in our analysis in Section~\ref{sec:periodic} (and a few periodic dCs) show signs of non-periodic variability, as well as secular, long-term trends. These non-periodic but variable dCs are of interest and may be signs of flaring, variable obscuration, or perhaps  accretion onto the WD companion. They warrant further investigation, but we do not discuss them further in this paper. 

The light curves that show long-term trends of brightening or dimming on 100s of days timescales cause the mean magnitude to vary over the entire time span of the light curve. This variable mean magnitude can cause issues with our period search. Therefore, we removed these long term trends by fitting out a third-order polynomial to the raw light curve.

\section{Periodic Variability}\label{sec:periodic}

For each light curve, we searched for periodic signals down to a minimum period of 0.1\,d using the the Lomb-Scargle periodogram \citep[LS;][]{Lomb1976, Scargle1982}. \added{We used the Astropy \citep{astropy2} implementation of the LS algorithm \citep{astropyLS1, astropyLS2}.} We selected the highest peak, and if this peak corresponds to an observational alias (1d, 29.5d, 1yr, etc.) or a harmonic of one of these aliases (1/2, 1/3, 1/4, 1/5, 2, 3, 4, 5), we removed that signal from the light curve and recalculated the periodogram until the highest-power frequency was not an alias (we counted a frequency not as an alias if it was more than 150 frequency bins away from the pure alias frequency, i.e. more than 0.005~d$^{-1}$ away from an alias).

For the highest remaining peak, we calculated the false-alarm probability \citep{Vanderplas2018}. We required that $\log{\left( \textrm{FAP} \right)} \leq -5$ in at least one filter for us to select a specific dC as a periodic candidate, more conservative than e.g., the $\log{\left( \textrm{FAP} \right)} \leq -3$ used in the recent ZTF periodic variable catalog of \citet{ChenX2020}.

\begin{figure*}
\centering
\plotone{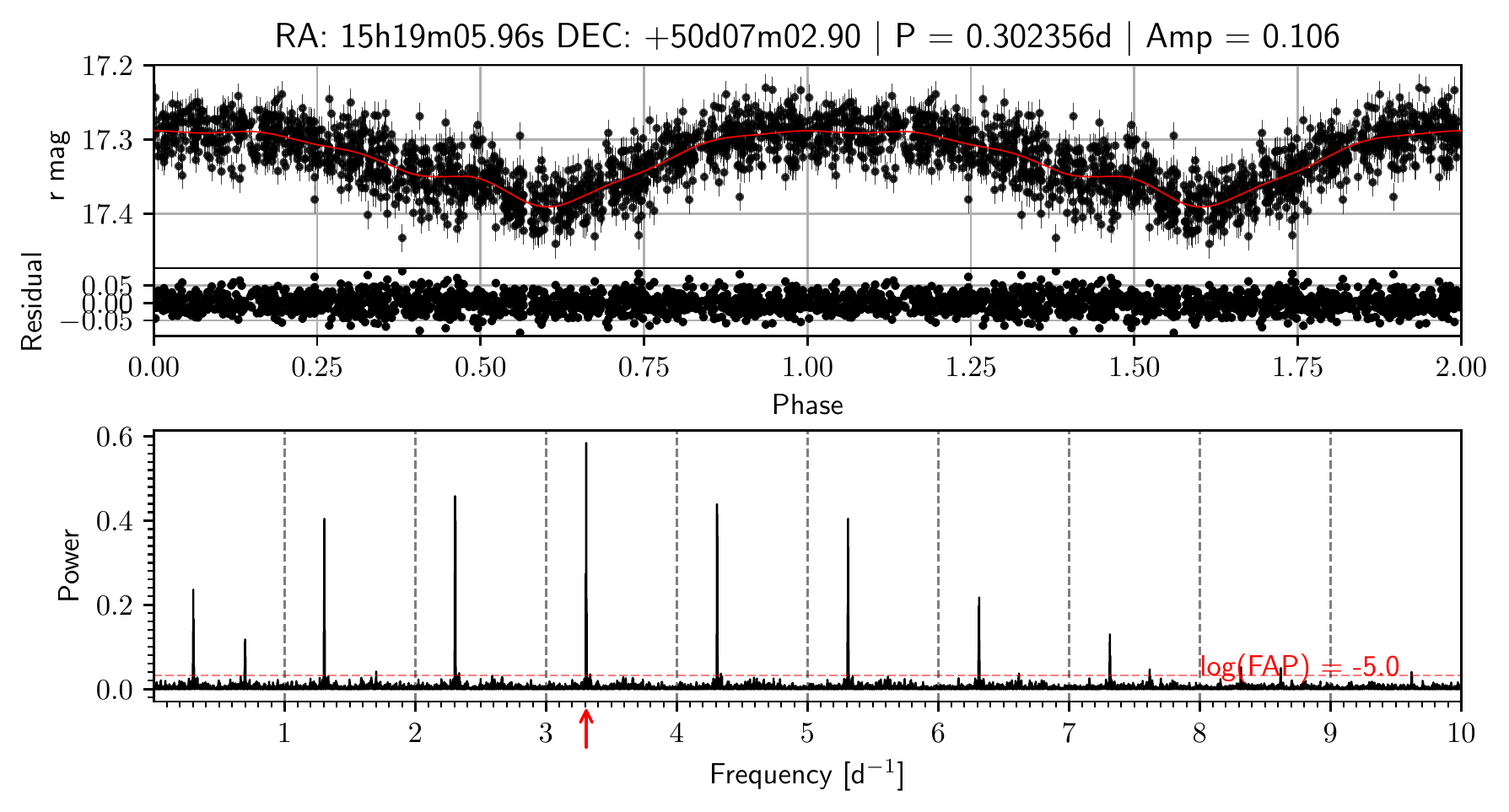}
\caption{Phased light curves and power spectra for all periodic dC candidates. This example light curve and power spectrum is for the dC \object{SDSS J151905.96+500702.9}. This dC shows a clear and strong signal at 3.3\,d$^{-1}$ (with 1-d aliasing) that stands out above the low noise background in the power spectrum. The red horizontal line represents the peak height needed for a signal to meet our $\log{\left( \textrm{FAP} \right)} \leq -5$ criterion. Grey vertical dashed lines mark the 1-d aliases caused by gaps in data collection, and the best period from our analysis is marked by an arrow. The highest significance peak is used to fold the observed light curve, yielding the phase-folded light curve in the top panel. Each light curve is plotted twice in phase to clearly show the periodic variability. The data are shown as the black scatter points with their respective error bars, and the best fitting model (see Section~\ref{sec:periodic}) is marked by the red solid line. The residuals are shown below the light curve. The complete figure set (35 figures) is available in the online journal.}
\label{fig:powerspec_compare}
\end{figure*}

For the dCs which have light curves selected as periodic candidates, we checked for any possible harmonic confusion in the found period. For each dC, in each filter, we plot a power spectrum from the LS analysis. This is used to determine how strong the highest-power frequency is compared to the $\log{\left( \textrm{FAP} \right)}$ limit and the background peaks. Figure~\ref{fig:powerspec_compare} shows an example power spectrum for an object with a very strong periodic signal and shows clear peaks (with 1-d aliasing) above the background, and the resulting phased light curve. The complete figure set (90 figures) is available in the online journal.

\input{LC_figset.tex}

In some cases, the strongest peaks were aliases, typically harmonics of 1 month, that overwhelmed the power spectrum. For these dCs, we inspected each power spectrum in conjunction with phased light curves. If another non-alias peak (i.e., with a frequency more than 0.005~d$^{-1}$ away from an alias) was found in the power spectrum meeting our FAP limit, that new peak was selected as the period for that dC. If no non-alias peaks could be found, the dC candidate was removed from our sample. 

Some dCs show strong periodic signals in one filter, but do not reach our FAP limit in the other available filters. For these dCs, if one filter has a period that meets our FAP limit and that period is visible in the other filter, we include that second filter even if its FAP does not meet our limit. This makes it possible for some dCs to have a $\log{\left( \textrm{FAP} \right)} \geq -5$ in a filter if they have $\log{\left( \textrm{FAP} \right)} \leq -5$ in another filter.

For all periodic dC candidates selected after inspection of their power spectra, we plotted phased light curves folded on the highest selected peak period. In addition, we plotted 8 different harmonics of that period (1/2, 1/3, 1/4, 1/5, 2, 3, 4, 5) to check for aliases caused by gaps in the observational coverage. Using this plot, we calculated model-fit statistics ($\chi^2$) and selected which period harmonic has the best model fit. We used the best period to phase the light curve, to which we fit the final periodic model.




 Our best-fit models were computed using the automatic Fourier decomposition (AFD) method, as detailed in \citet{Torrealba2015}. We set an upper limit on the number of Fourier series terms of $n_{max} = 6$ to reduce over-fitting. No significant non-harmonic terms were included; though one dC, LAMOST~J062558.33+023019.4, showed different peaks in its power spectrum between the g and r filters with the second highest peak in each filter being the highest peak in the other. The best AFD model was used to calculate the amplitude and epoch of brightest time ($t_0$) for each light curve.  We removed any dC for which the folded light curve shows no clear periodic signal or for which the amplitude of the variability was less than 0.005\,mag. \added{For dCs with multiple filters, we use the period from the filter with the strongest detection as the selected period and force the fits in the other filters to this fixed period. Some filters may not have a clear detection from the signal found in another filter, resulting in model fits with large errors for that filter. Figure Set 1 contains the folded light curves, models, and power spectra for all the periodic dC candidates.}

Table~\ref{tab:periodic_prop} contains the properties for this final periodic dC sample. We estimated errors for the best found period using a Markov-Chain Monte Carlo (MCMC) method. For each dC, in each filter with a detected period, we started 50 MCMC walkers in a Gaussian around the detected period. We sampled the walkers for 10,000 steps each, at each step using the phase dispersion minimization technique \citep{Stellingwerf1978} to calculate the likelihood at each walker position. We used the 1$\sigma$ of the marginalized period distribution as the photometric period error for that dC.  However this is only a statistical error, and does not account for the possibility that we have selected an alias rather than the true period. 

Our final dC sample contains \replaced{24}{34} individual dCs that are periodic in at least one ZTF filter. Given the wide initial orbits necessary for progenitor dC systems to avoid truncation of the TP-AGB phase before enough C-rich material can be transferred, it is remarkable that \replaced{19 (80\%)}{19 (56\%)} of these dCs have periods $< 1$\,d \added{(and 28 (82\%) of these dCs have periods $< 2$\,d)}, indicating they should have experienced a common-envelope (CE) phase. The likely origins of the variability in these dCs include spot rotation on the dC or tidal distortion of the dC atmosphere from being in a close orbit with a WD. Since many of these dCs have short periods, we assume that these systems would have experienced a CE phase and have circularized and synchronized \citep{Hurley2002}. However, if the light curve variability is from the dC being tidally distorted, our detected period would be half the orbital period (even with 2$\times$ longer true orbital periods, these systems should still have experienced a CE phase). 

\added{In addition, the 1\,d aliasing caused by the observing window function causes peaks at frequencies of $\pm 1$\,d$^{-1}$. These alias peaks can also meet our $\log{\left( \textrm{FAP} \right)}$ limit (see Figure \ref{fig:powerspec_compare}), and while we take the highest significance peak from the filter which produces the best fitting model (via a $\chi^2$ fit) there is a possibility this is the wrong period. This can only be solved by either low cadence photometry or confirming the photometric period with RV follow-up. For example, the dC \object{SBSS 1310+561} has $\pm 1$\,d$^{-1}$ aliases the meet our $\log{\left( \textrm{FAP} \right)}$ limit. In our initial search using ZTF DR3 data, the $g$ and $r$ filters had different highest peaks, with the best period in the $g$ filter being $\sim$5.18\,d and the best period in the  r filter being $\sim$0.838\,d. These are separated exactly by the $1$\,d$^{-1}$ aliasing of the window function, with the $r$ filter providing a better model fit. However, using the newest (and larger) ZTF DR5 data set results in both the $g$ and $r$ filters having the same highest peak, at $5.1878 \pm 0.0012$\,d. \citet{Whitehouse2021} confirmed this as the period with their RV observations of this dC.}

\added{The recent work by \citet{Whitehouse2021} included modeling of light curves for their sample of periodic dCs. They examined whether spot rotation, tidal distortion or irradiation by a hot WD companion could be the source of the photometric variability in their dC sample. They found that for periods near and longer than $1$\,d, both tidal distortion and irradiation are reduced to levels that would not be detectable in the light curves. While tidal distortion could be detectable for our shortest period dCs in this paper ($0.1 < $ P $< 0.2$\,d), the predicted amplitudes at these periods from \citet{Whitehouse2021} are larger than those we find in our sample (as was the case for their sample). The irradiation modeling of \citet{Whitehouse2021} (which assumes a WD temperature from 30,000\,K to 20,000\,K) predicts amplitudes large enough to be detected. However, the majority of our dC amplitudes are smaller than predicted by the models, suggesting irradiation is not the source of variability for most of our dCs. Additionally, the majority of the dCs in our sample are mid- and late-type dCs \citep[see][]{Roulston2019} which do not have a visible WD in their optical spectra. This sets a limit that for these types, that WD must be cooler than about 10,000\,K, reducing the irradiation effects below our detection limits. Only the composite dC+WD systems which have a hot WD (like \object{SDSS J151905.96+500702.9}) may have detectable irradiation effects. This leaves spot rotation as the most likely source of variability in our periodic dC sample. However, the origin of the variability in these dCs is not truly confirmed without comparison to spectroscopic RV follow-up.} \deleted{Tidal distortion should introduce a non-sinusoidal shape to the light curves. However, there is little evidence of this in our sample, lending some evidence that most of the detected variability is probably from spot rotation.  Therefore, the origin of the variability in these dCs is not yet fully known without comparison to spectroscopic RV follow-up.}

\added{Finally, as the ZTF survey continues to accure more data we expect to find more photometrically variable dCs from the current sample of known dCs. However, even given a favorable inclination (say $i > 85^\circ$) the ZTF errors are too large for the detection of an eclipse of a cool WD in these systems. Using our estimated dC radii and luminosities and assuming a WD companion with a standard mass of $0.6$M$_\odot$ and temperature of 7000K (we expect to see the WD component in the optical spectra if it is any hotter than this) we would only expect an average primary eclipse depth of 0.005 mag. This is below our detection threshold with ZTF, with our dCs having median errors of 0.019 mag, compared to their median amplitude of 0.059\,mag. The Vera Rubin Observatory's LSST survey \citep{LSST} is expected to have errors of approximately 0.005 mag for a point source with r $= 19.0$, which may allow for the detection of dC eclipses. However, the majority of known dCs reside outside the LSST footprint, with 17\% below the declination cut of of $\delta = +2^\circ$ \citep{LSST}. Detection and characterization of the first eclipsing dC system will likely require dedicated observations with high cadence and low photometric noise.}

\begin{longrotatetable}
\begin{deluxetable}{LcccDDDDDDDDDDDcD}
\tabletypesize{\scriptsize}
\centerwidetable
\movetabledown=0.4in
\tablecaption{Periodic dC Light Curve Properties}
\tablehead{\colhead{Index$^\tablenotemark{a}$} & \colhead{R.A.} & \colhead{Decl.} & \colhead{Filter} & \twocolhead{$N_{\textrm{good}}$} & \twocolhead{$N_{\textrm{rejects}}$} & \twocolhead{$<\textrm{mag}>$} & \twocolhead{$<\sigma_{\rm{mag}}>$} & \twocolhead{P} & \twocolhead{$\sigma_{\textrm{P}}$} & \twocolhead{$\log_{10}{\textrm{FAP}}$} & \twocolhead{Amp} & \twocolhead{$\sigma_{\textrm{Amp}}$} & \twocolhead{t$_0$} & \twocolhead{$\sigma_{t_0}$} &  \colhead{$N_{\textrm{terms}}$} & \twocolhead{$\chi^2_{\nu,\textrm{fit}}$} \\ \colhead{} & \colhead{(J2016.0)} & \colhead{(J2016.0)} & \colhead{}  & \twocolhead{} & \twocolhead{} & \twocolhead{[mag]} & \twocolhead{[mag]} & \twocolhead{[d]} & \twocolhead{[d]} & \twocolhead{} & \twocolhead{[mag]} & \twocolhead{[mag]} & \twocolhead{[d]} & \twocolhead{[d]} &  \colhead{} & \twocolhead{}}
\decimals
\startdata
1^*        & 00h47m06.76s & +00d07m48.80s & g & 183  & 0  & 19.81  & 0.12  & 12.614   & 0.010    & $>$-1.0 & 0.044  & 0.051  & 59217.686   & 0.022   & 1 & 1.18 \\
2          & 00h47m06.76s & +00d07m48.80s & r & 243  & 1  & 18.352 & 0.043 & 12.614   & 0.010    & -8.9    & 0.058  & 0.015  & 59217.383   & 0.022   & 1 & 0.87 \\
3^*        & 01h31m19.05s & +37d20m25.30s & g & 179  & 2  & 18.555 & 0.037 & 1.376804 & 0.000079 & $>$-1.0 & 0.055  & 0.015  & 59066.1913  & 0.0014  & 1 & 2.42 \\
4^*        & 01h31m19.05s & +37d20m25.30s & i & 27   & 0  & 16.906 & 0.020 & 1.376804 & 0.000079 & $>$-1.0 & 0.026  & 0.022  & 58746.5621  & 0.0014  & 1 & 1.73 \\
5          & 01h31m19.05s & +37d20m25.30s & r & 236  & 1  & 17.266 & 0.017 & 1.376804 & 0.000079 & -21.1   & 0.0553 & 0.0063 & 59067.5006  & 0.0014  & 1 & 2.17 \\
6          & 02h35m30.65s & +02d25m18.58s & g & 209  & 1  & 17.993 & 0.031 & 1.65926  & 0.00019  & -3.3    & 0.050  & 0.012  & 59232.6027  & 0.0017  & 1 & 2.26 \\
7          & 02h35m30.65s & +02d25m18.58s & r & 229  & 1  & 16.673 & 0.015 & 1.65926  & 0.00019  & -10.1   & 0.0416 & 0.0080 & 59232.6126  & 0.0017  & 2 & 2.37 \\
8          & 02h54m14.24s & +26d21m54.19s & g & 255  & 2  & 15.598 & 0.012 & 0.190058 & 0.000018 & -11.5   & 0.0351 & 0.0043 & 59232.01892 & 0.00019 & 1 & 9.73 \\
9          & 02h54m14.24s & +26d21m54.19s & r & 262  & 3  & 14.529 & 0.010 & 0.190058 & 0.000018 & -4.7    & 0.0228 & 0.0033 & 59232.00485 & 0.00019 & 1 & 10.03 \\
10         & 04h16m05.11s & +50d28m28.52s & g & 293  & 0  & 14.375 & 0.013 & 6.8083   & 0.0018   & -7.5    & 0.0212 & 0.0044 & 59228.4734  & 0.0070  & 1 & 1.66 \\
11         & 04h16m05.11s & +50d28m28.52s & r & 317  & 1  & 13.601 & 0.012 & 6.8083   & 0.0018   & -6.6    & 0.0163 & 0.0054 & 59227.6972  & 0.0070  & 2 & 1.20 \\
12^*       & 05h02m40.82s & +40d23m23.59s & g & 72   & 0  & 18.414 & 0.037 & 4.42791  & 0.00096  & $>$-1.0 & 0.060  & 0.025  & 58750.1829  & 0.0045  & 1 & 2.32 \\
13         & 05h02m40.82s & +40d23m23.59s & r & 294  & 0  & 17.124 & 0.019 & 4.42791  & 0.00096  & -32.6   & 0.0646 & 0.0063 & 59233.0912  & 0.0045  & 1 & 1.81 \\
14         & 06h25m58.34s & +02d30m19.43s & g & 157  & 0  & 14.589 & 0.015 & 7.6080   & 0.0014   & -13.6   & 0.0660 & 0.0067 & 59227.6546  & 0.0077  & 1 & 3.51 \\
15         & 06h25m58.34s & +02d30m19.43s & r & 176  & 2  & 13.885 & 0.010 & 7.6080   & 0.0014   & -16.6   & 0.0497 & 0.0043 & 59227.4415  & 0.0077  & 1 & 7.68 \\
16         & 07h44m47.66s & +51d38m31.76s & g & 220  & 2  & 17.506 & 0.022 & 1.534684 & 0.000071 & -1.4    & 0.0435 & 0.0084 & 59230.5473  & 0.0015  & 1 & 3.88 \\
17         & 07h44m47.66s & +51d38m31.76s & r & 279  & 1  & 16.122 & 0.011 & 1.534684 & 0.000071 & -20.5   & 0.0426 & 0.0055 & 59232.0743  & 0.0015  & 2 & 3.62 \\
18^*       & 08h11m57.14s & +14d35m33.00s & g & 116  & 2  & 16.035 & 0.014 & 0.750413 & 0.000041 & $>$-1.0 & 0.0150 & 0.0069 & 59230.04781 & 0.00075 & 1 & 2.42 \\
19         & 08h11m57.14s & +14d35m33.00s & r & 210  & 0  & 15.742 & 0.013 & 0.750413 & 0.000041 & -14.3   & 0.0541 & 0.0059 & 59232.20075 & 0.00075 & 1 & 3.11 \\
20^*       & 09h14m58.08s & +21d56m39.65s & g & 108  & 0  & 16.839 & 0.016 & 1.23573  & 0.00014  & $>$-1.0 & 0.0366 & 0.0090 & 59231.8032  & 0.0012  & 1 & 7.97 \\
21         & 09h14m58.08s & +21d56m39.65s & r & 202  & 2  & 15.330 & 0.010 & 1.23573  & 0.00014  & -7.1    & 0.0471 & 0.0045 & 59231.8045  & 0.0012  & 1 & 7.59 \\
22^{*\dag} & 09h33m24.58s & -00d31m44.07s & g & 107  & 0  & 15.308 & 0.013 & 1.15693  & 0.00014  & $>$-1.0 & 0.0091 & 0.0072 & 59231.8412  & 0.0012  & 1 & 3.66 \\
23^\dag    & 09h33m24.58s & -00d31m44.07s & r & 268  & 3  & 13.989 & 0.011 & 1.15693  & 0.00014  & -14.2   & 0.0196 & 0.0036 & 59231.4432  & 0.0012  & 1 & 1.75 \\
24^*       & 09h40m26.28s & +36d25m48.81s & g & 261  & 1  & 19.79  & 0.12  & 1.9573   & 0.0012   & $>$-1.0 & 0.070  & 0.040  & 59228.7471  & 0.0023  & 1 & 1.32 \\
25^*       & 09h40m26.28s & +36d25m48.81s & i & 22   & 0  & 17.740 & 0.038 & 1.9573   & 0.0012   & $>$-1.0 & 0.004  & 0.044  & 58627.9758  & 0.0023  & 1 & 1.50 \\
26         & 09h40m26.28s & +36d25m48.81s & r & 581  & 3  & 18.241 & 0.038 & 1.9573   & 0.0012   & -14.6   & 0.044  & 0.0082 & 59230.8844  & 0.0023  & 1 & 1.35 \\
27^*       & 12h02m46.01s & +54d19m29.24s & g & 249  & 0  & 20.68  & 0.21  & 1.15516  & 0.00024  & $>$-1.0 & 0.045  & 0.074  & 59231.1841  & 0.0012  & 1 & 1.14 \\
28^*       & 12h02m46.01s & +54d19m29.24s & i & 22   & 0  & 18.397 & 0.053 & 1.15516  & 0.00024  & $>$-1.0 & 0.038  & 0.065  & 58651.6814  & 0.0012  & 1 & 0.92 \\
29         & 12h02m46.01s & +54d19m29.24s & r & 460  & 0  & 18.917 & 0.073 & 1.15516  & 0.00024  & -5.9    & 0.056  & 0.019  & 59231.3031  & 0.0012  & 1 & 0.71 \\
30         & 12h08m53.35s & -00d08m47.99s & g & 59   & 0  & 19.66  & 0.10  & 0.350882 & 0.000012 & -5.7    & 0.214  & 0.076  & 59234.43466 & 0.00035 & 1 & 0.63 \\
31^*       & 12h08m53.35s & -00d08m47.99s & r & 92   & 1  & 18.701 & 0.063 & 0.350882 & 0.000012 & $>$-1.0 & 0.019  & 0.037  & 59231.29497 & 0.00035 & 1 & 1.62 \\
32^*       & 12h10m06.99s & +58d43m18.34s & g & 441  & 1  & 17.998 & 0.038 & 0.183532 & 0.000010 & $>$-1.0 & 0.005  & 0.011  & 59232.46412 & 0.00018 & 1 & 1.79 \\
33^*       & 12h10m06.99s & +58d43m18.34s & i & 25   & 0  & 16.577 & 0.014 & 0.183532 & 0.000010 & $>$-1.0 & 0.074  & 0.020  & 58652.08785 & 0.00018 & 1 & 13.89 \\
34         & 12h10m06.99s & +58d43m18.34s & r & 616  & 5  & 16.868 & 0.017 & 0.183532 & 0.000010 & -7.6    & 0.0366 & 0.0055 & 59232.47201 & 0.00018 & 2 & 5.62 \\
35^{*\dag} & 12h23m57.62s & +55d01m51.43s & g & 856  & 3  & 19.074 & 0.078 & 0.336288 & 0.000032 & $>$-1.0 & 0.072  & 0.036  & 59231.38908 & 0.00034 & 6 & 1.26 \\
36^{*\dag} & 12h23m57.62s & +55d01m51.43s & i & 53   & 0  & 16.977 & 0.020 & 0.336288 & 0.000032 & $>$-1.0 & 0.034  & 0.026  & 58675.06583 & 0.00034 & 2 & 1.00 \\
37^\dag    & 12h23m57.62s & +55d01m51.43s & r & 1001 & 4  & 17.399 & 0.022 & 0.336288 & 0.000032 & -16.2   & 0.0257 & 0.0057 & 59231.46542 & 0.00034 & 2 & 1.20 \\
38         & 12h30m45.52s & +41d09m43.45s & g & 682  & 4  & 18.362 & 0.041 & 0.882519 & 0.000020 & -14.6   & 0.0545 & 0.0089 & 59231.23169 & 0.00088 & 1 & 1.76 \\
39^*       & 12h30m45.52s & +41d09m43.45s & i & 46   & 0  & 15.788 & 0.015 & 0.882519 & 0.000020 & $>$-1.0 & 0.016  & 0.012  & 58647.83545 & 0.00088 & 1 & 1.06 \\
40         & 12h30m45.52s & +41d09m43.45s & r & 624  & 0  & 16.566 & 0.019 & 0.882519 & 0.000020 & -57.7   & 0.0417 & 0.0042 & 59229.43135 & 0.00088 & 1 & 1.10 \\
41^*       & 13h03m59.18s & +05d09m38.62s & g & 105  & 0  & 18.361 & 0.039 & 1.84149  & 0.00014  & $>$-1.0 & 0.048  & 0.021  & 59229.5733  & 0.0018  & 1 & 2.12 \\
42^*       & 13h03m59.18s & +05d09m38.62s & i & 23   & 0  & 16.757 & 0.019 & 1.84149  & 0.00014  & $>$-1.0 & 0.041  & 0.027  & 58651.6711  & 0.0018  & 1 & 3.87 \\
43         & 13h03m59.18s & +05d09m38.62s & r & 124  & 0  & 17.101 & 0.020 & 1.84149  & 0.00014  & -10.5   & 0.070  & 0.015  & 59231.6633  & 0.0018  & 2 & 2.10 \\
44         & 13h12m42.27s & +55d55m54.84s & g & 403  & 4  & 15.845 & 0.014 & 5.1878   & 0.0012   & -8.1    & 0.0439 & 0.0068 & 59229.5411  & 0.0053  & 3 & 6.00 \\
45         & 13h12m42.27s & +55d55m54.84s & i & 35   & 0  & 13.560 & 0.014 & 5.1878   & 0.0012   & -1.1    & 0.034  & 0.016  & 58659.3606  & 0.0053  & 1 & 1.19 \\
46         & 13h12m42.27s & +55d55m54.84s & r & 402  & 2  & 14.121 & 0.012 & 5.1878   & 0.0012   & -28.1   & 0.0343 & 0.0035 & 59230.0028  & 0.0053  & 1 & 2.50 \\
47         & 13h31m23.61s & +48d26m24.37s & g & 314  & 4  & 20.276 & 0.154 & 0.203571 & 0.000043 & -5.5    & 0.43   & 0.12   & 59231.42815 & 0.00021 & 6 & 2.13 \\
48^*       & 13h31m23.61s & +48d26m24.37s & i & 29   & 0  & 18.194 & 0.040 & 0.203571 & 0.000043 & $>$-1.0 & 0.020  & 0.041  & 58661.03320 & 0.00021 & 1 & 0.55 \\
49^*       & 13h31m23.61s & +48d26m24.37s & r & 437  & 4  & 18.640 & 0.045 & 0.203571 & 0.000043 & $>$-1.0 & 0.008  & 0.012  & 59233.29307 & 0.00021 & 1 & 1.14 \\
50^*       & 14h09m53.08s & -06d11m41.71s & g & 184  & 1  & 15.275 & 0.014 & 0.319873 & 0.000014 & $>$-1.0 & 0.0114 & 0.0060 & 59231.23975 & 0.00032 & 1 & 3.87 \\
51^*       & 14h09m53.08s & -06d11m41.71s & i & 20   & 0  & 13.995 & 0.014 & 0.319873 & 0.000014 & $>$-1.0 & 0.011  & 0.018  & 58652.94999 & 0.00032 & 1 & 1.88 \\
52         & 14h09m53.08s & -06d11m41.71s & r & 260  & 0  & 14.268 & 0.013 & 0.319873 & 0.000014 & -5.6    & 0.0208 & 0.0067 & 59232.20001 & 0.00032 & 2 & 1.62 \\
53         & 14h15m15.24s & +51d41m28.01s & g & 322  & 0  & 20.64  & 0.20  & 0.272819 & 0.000018 & -5.2    & 0.197  & 0.062  & 59231.22128 & 0.00027 & 1 & 1.06 \\
54^*       & 14h15m15.24s & +51d41m28.01s & i & 35   & 0  & 19.47  & 0.11  & 0.272819 & 0.000018 & $>$-1.0 & 0.040  & 0.098  & 58711.95424 & 0.00027 & 1 & 1.48 \\
55^*       & 14h15m15.24s & +51d41m28.01s & r & 539  & 1  & 19.67  & 0.11  & 0.272819 & 0.000018 & $>$-1.0 & 0.026  & 0.026  & 59231.27530 & 0.00027 & 1 & 0.69 \\
56^*       & 15h11m44.58s & +38d59m10.46s & g & 509  & 2  & 18.768 & 0.055 & 0.335548 & 0.000082 & $>$-1.0 & 0.009  & 0.014  & 59202.32548 & 0.00034 & 1 & 2.09 \\
57^*       & 15h11m44.58s & +38d59m10.46s & i & 45   & 0  & 16.702 & 0.017 & 0.335548 & 0.000082 & $>$-1.0 & 0.010  & 0.016  & 58733.05657 & 0.00034 & 1 & 1.18 \\
58         & 15h11m44.58s & +38d59m10.46s & r & 570  & 1  & 17.217 & 0.020 & 0.335548 & 0.000082 & -5.2    & 0.0179 & 0.0066 & 59202.22985 & 0.00034 & 2 & 1.23 \\
59         & 15h15m42.72s & +52d01m45.47s & g & 529  & 7  & 18.661 & 0.044 & 0.332473 & 0.000097 & -5.0    & 0.067  & 0.019  & 59231.21797 & 0.00033 & 3 & 2.37 \\
60^*       & 15h15m42.72s & +52d01m45.47s & r & 527  & 6  & 17.253 & 0.018 & 0.332473 & 0.000097 & $>$-1.0 & 0.0092 & 0.0046 & 59231.48893 & 0.00033 & 1 & 3.45 \\
61         & 15h19m05.93s & +50d07m03.14s & g & 1279 & 5  & 17.609 & 0.023 & 0.302356 & 0.000021 & -10.2   & 0.0184 & 0.0052 & 59231.41069 & 0.00030 & 2 & 1.18 \\
62         & 15h19m05.93s & +50d07m03.14s & i & 104  & 0  & 17.005 & 0.021 & 0.302356 & 0.000021 & -11.6   & 0.132  & 0.017  & 58733.12921 & 0.00030 & 2 & 3.19 \\
63         & 15h19m05.93s & +50d07m03.14s & r & 1281 & 6  & 17.325 & 0.018 & 0.302356 & 0.000021 & -238.8  & 0.1061 & 0.0072 & 59233.23027 & 0.00030 & 6 & 2.03 \\
64         & 15h24m34.12s & +44d49m55.84s & g & 236  & 1  & 20.85  & 0.19  & 0.251714 & 0.000012 & -5.1    & 0.36   & 0.12   & 59231.38694 & 0.00025 & 3 & 1.22 \\
65^*       & 15h24m34.12s & +44d49m55.84s & i & 23   & 0  & 18.693 & 0.065 & 0.251714 & 0.000012 & $>$-1.0 & 0.009  & 0.079  & 58669.14143 & 0.00025 & 1 & 0.74 \\
66^*       & 15h24m34.12s & +44d49m55.84s & r & 448  & 1  & 19.206 & 0.067 & 0.251714 & 0.000012 & $>$-1.0 & 0.025  & 0.018  & 59233.37775 & 0.00025 & 1 & 1.16 \\
67^*       & 15h25m04.49s & +32d25m10.90s & g & 412  & 3  & 21.06  & 0.20  & 0.13712  & 0.000013 & $>$-1.0 & 0.27   & 0.13   & 59232.50195 & 0.00014 & 5 & 1.96 \\
68^*       & 15h25m04.49s & +32d25m10.90s & i & 97   & 0  & 19.198 & 0.083 & 0.13712  & 0.000013 & $>$-1.0 & 0.030  & 0.045  & 58733.08198 & 0.00014 & 1 & 2.62 \\
69         & 15h25m04.49s & +32d25m10.90s & r & 1001 & 3  & 19.557 & 0.088 & 0.13712  & 0.000013 & -5.4    & 0.070  & 0.016  & 59232.43462 & 0.00014 & 1 & 2.11 \\
70^*       & 15h30m59.26s & +45d12m00.33s & g & 504  & 1  & 18.095 & 0.032 & 13.587   & 0.011    & $>$-1.0 & 0.0389 & 0.0082 & 59090.790   & 0.018   & 1 & 2.65 \\
71^*       & 15h30m59.26s & +45d12m00.33s & i & 36   & 0  & 16.059 & 0.014 & 13.587   & 0.011    & $>$-1.0 & 0.027  & 0.014  & 58721.469   & 0.018   & 1 & 1.61 \\
72         & 15h30m59.26s & +45d12m00.33s & r & 528  & 4  & 16.579 & 0.014 & 13.587   & 0.011    & -25.6   & 0.0295 & 0.0036 & 59062.217   & 0.018   & 1 & 1.87 \\
73^*       & 15h35m32.92s & +01d10m16.22s & g & 31   & 0  & 21.02  & 0.19  & 0.173866 & 0.000037 & $>$-1.0 & 0.14   & 0.21   & 59038.20950 & 0.00018 & 1 & 0.80 \\
74^*       & 15h35m32.92s & +01d10m16.22s & i & 23   & 0  & 18.514 & 0.053 & 0.173866 & 0.000037 & $>$-1.0 & 0.019  & 0.062  & 58667.09530 & 0.00018 & 1 & 1.53 \\
75         & 15h35m32.92s & +01d10m16.22s & r & 177  & 0  & 19.061 & 0.085 & 0.173866 & 0.000037 & -8.0    & 0.233  & 0.037  & 59231.42104 & 0.00018 & 1 & 2.33 \\
76^*       & 16h37m18.63s & +27d40m26.63s & g & 544  & 4  & 19.295 & 0.064 & 1.22790  & 0.00010  & $>$-1.0 & 0.048  & 0.022  & 59232.4009  & 0.0012  & 2 & 1.74 \\
77^*       & 16h37m18.63s & +27d40m26.63s & i & 48   & 0  & 16.759 & 0.016 & 1.22790  & 0.00010  & $>$-1.0 & 0.035  & 0.025  & 58732.2916  & 0.0012  & 3 & 1.21 \\
78         & 16h37m18.63s & +27d40m26.63s & r & 578  & 1  & 17.424 & 0.018 & 1.22790  & 0.00010  & -7.5    & 0.0191 & 0.0043 & 59232.1419  & 0.0012  & 1 & 1.34 \\
79^*       & 16h59m02.30s & +25d05m49.00s & g & 250  & 2  & 21.08  & 0.19  & 0.287694 & 0.000030 & $>$-1.0 & 0.242  & 0.097  & 59232.53484 & 0.00029 & 2 & 2.65 \\
80^*       & 16h59m02.30s & +25d05m49.00s & i & 54   & 0  & 18.769 & 0.056 & 0.287694 & 0.000030 & $>$-1.0 & 0.049  & 0.043  & 58751.08757 & 0.00029 & 1 & 1.52 \\
81         & 16h59m02.30s & +25d05m49.00s & r & 628  & 6  & 19.358 & 0.064 & 0.287694 & 0.000030 & -5.7    & 0.083  & 0.025  & 59232.43645 & 0.00029 & 3 & 2.73 \\
82^*{\dag} & 19h23m55.93s & +44d58m32.20s & g & 719  & 10 & 17.286 & 0.019 & 0.146029 & 0.000013 & $>$-1.0 & 0.0058 & 0.0039 & 59194.06850 & 0.00015 & 1 & 2.82 \\
83^{*\dag} & 19h23m55.93s & +44d58m32.20s & i & 56   & 0  & 16.009 & 0.014 & 0.146029 & 0.000013 & $>$-1.0 & 0.011  & 0.011  & 58748.21787 & 0.00015 & 1 & 1.74 \\
84^\dag    & 19h23m55.93s & +44d58m32.20s & r & 1077 & 2  & 16.285 & 0.013 & 0.146029 & 0.000013 & -15.3   & 0.0150 & 0.0032 & 59194.01389 & 0.00015 & 2 & 1.68 \\
85         & 22h08m10.01s & +25d17m30.17s & g & 218  & 2  & 15.724 & 0.014 & 0.422469 & 0.000014 & -20.4   & 0.0454 & 0.0055 & 59223.02886 & 0.00042 & 1 & 20.78 \\
86         & 22h08m10.01s & +25d17m30.17s & i & 48   & 0  & 14.122 & 0.014 & 0.422469 & 0.000014 & -3.0    & 0.032  & 0.012  & 58750.73006 & 0.00042 & 1 & 1.19 \\
87         & 22h08m10.01s & +25d17m30.17s & r & 291  & 1  & 14.548 & 0.014 & 0.422469 & 0.000014 & -40.7   & 0.0541 & 0.0043 & 59228.94681 & 0.00042 & 1 & 2.20 \\
88         & 23h41m30.74s & +15d19m43.20s & g & 507  & 5  & 18.338 & 0.036 & 0.134337 & 0.000019 & -97.1   & 0.1437 & 0.0088 & 59233.04549 & 0.00013 & 1 & 1.83 \\
89         & 23h41m30.74s & +15d19m43.20s & i & 81   & 0  & 17.513 & 0.028 & 0.134337 & 0.000019 & -8.9    & 0.094  & 0.018  & 58751.30804 & 0.00013 & 1 & 1.36 \\
90         & 23h41m30.74s & +15d19m43.20s & r & 541  & 2  & 17.687 & 0.024 & 0.134337 & 0.000019 & -121.7  & 0.1205 & 0.0059 & 59233.04603 & 0.00013 & 1 & 1.62
\enddata
\tablenotetext{a}{\footnotesize Rows marked with an $^*$ have a light curve with no detectable variability at the given period in this filter. The model fit for this filter is unreliable. dCs marked with \dag, while meeting all our criteria to be included, have suspect periods due to having more than 1\% of the periodogram above the power needed to have $\log{\left( \textrm{FAP} \right)} \leq -5$.}
\tablecomments{Periodic properties of dwarf carbon stars found in this paper. For each dC, we list light curve properties of the observed ZTF filter, the mean magnitude and mean magnitude error. We include the selected best period, the logarithm of the false-alarm-probability for that period, the amplitude of variability from the best fit model at that period, and the time of light curve maximum brightness. Finally, we include a few diagnostics including the number of terms in our model fit and the resulting reduced $\chi^2$ of the model.} 
\end{deluxetable}
\end{longrotatetable}
\label{tab:periodic_prop}
\section{Spectroscopic Follow-up}\label{sec:rvs}

To constrain the origins of the photometric variability we have begun spectroscopic follow-up of the periodic dCs discovered here. We report spectroscopic follow-up for four of these dCs: \object{SDSS J151905.96+500702.9}, \object{SDSS J123045.53+410943.8}, \object{LAMOST J062558.33+023019.4} (referenced further on as J1519, J1230, J0625 respectively) and \object{SBSS 1310+561}.

\subsection{Spectroscopic Set-Up}\label{subsec:spec_setup}

The dCs J1519 and J1230 were observed with the Binospec spectrograph on the MMT telescope \citep{Fabricant2019}. For all observations, we used the 0.85\arcsec\ slit with the 600~l~mm$^{-1}$ grating centered on 7250\,\AA, giving coverage from $6000$\,\AA\ to $8000$\,\AA\ covering H$\alpha$ and the CN bands. The reduced spectra have a dispersion of 0.61\,\AA~pix$^{-1}$ with R$\approx$3590. All Binospec data were reduced using the standard Binospec reduction pipeline\footnote{\href{https://bitbucket.org/chil\_sai/binospec/wiki/Home}{https://bitbucket.org/chil\_sai/binospec/wiki/Home}} \citep{Binospec_reduc}.

The dC J0625 was observed with the Magellan Echellette \citep[MagE;][]{MagE} spectrograph on the Magellan Baade Telescope. All observations used the 0.85\arcsec\ slit and were reduced using the MagE reduction pipeline\footnote{\href{https://bitbucket.org/chil\_sai/mage-pipeline/src/master/}{https://bitbucket.org/chil\_sai/mage-pipeline/src/master/}} \citep{MagE_reduc}. The reduced spectra cover from about 3200\,\AA\ to 10000\,\AA\ with R$\approx$4500.

Observations for \object{SBSS 1310+561} were acquired at the 1.5m Fred Lawrence Whipple Observatory (FLWO) telescope with the FAST spectrograph \citep{FAST}  using the 600~l~mm$^{-1}$ grating and the 1.5\arcsec slit, which provides wavelength coverage from 6000\,\AA\ to 8000\,\AA\ at 1.5\,\AA\ spectral resolution. 

\subsection{SDSS J151905.96+500702.9}\label{subsec:229+50}

One of the more interesting dCs in our periodic sample, with photometric periodicity detected with highest significance, is SDSS J151905.96+500702.9 (also known as CBS~311; we use J1519 in the rest of this paper), a dC+DA spectroscopic composite binary. J1519 was discovered by \citet{Liebert1994} and has been studied on numerous occasions \citep{Farihi2010, Green2013, Whitehouse2018, Ashley2019, Roulston2019, Green2019}. However, this is the first reporting of its periodic variability.

J1519 ($r=17.3$\,mag) has four epochs of optical spectra in the SDSS, with the most recent spectrum shown in the top panel of Figure~\ref{fig:J1519_spec_rvHalpha}. The spectrum of J1519 shows a dC with a hot DA WD companion, as well as H$\alpha$ emission. \citet{Whitehouse2018} and \citet{Roulston2019} found RV variability using few-epoch spectroscopy with $\Delta RV_{\textrm{max}}$ of $46.8\pm15.8$~km~s$^{-1}$ and  $44\pm20$~km~s$^{-1}$, respectively. \citet{Farihi2010} conducted a study of WD–red dwarf systems, including J1519, using the \textit{Hubble Space Telescope}. They found J1519 to be unresolved, placing the constraint on its separation of $<10$\,au. 

\subsubsection{J1519 WD Model Fits}\label{subsec:wdprop}

Since J1519 is a spectroscopic dC+DA composite, we can fit WD model atmospheres to the WD component to fit \teff\ and \logg\ using the SDSS spectra. \citet{Bedard2021} fit WD models and found fit values of \replaced{31233$\pm$213\,K}{31230$\pm$210\,K} and 7.97$\pm$0.05 respectively.  

\citet{Farihi2010} found that spectroscopically fit WD parameters are often biased due to a cool companion. To update the fits of \citet{Bedard2021}, we performed our own model atmosphere fits to the DA component of J1519 using the synthetic WD model atmospheres of \citet{Levenhagen2017}. We first fit the late type dC (dCM) template of \citet{Roulston2020} to the SDSS spectrum of J1519 by finding the best-fit velocity, shifting the template, and then scaling it to the flux near H$\alpha$. We then removed the dC spectrum from the total spectrum, leaving just the WD component. We then fit the visible Balmer lines from H$\beta$ and blue-ward to the entire grid of WD model spectra. We interpolated the grid of WD model spectra to include half-steps in the model space. Our best-fitting model parameters for  \teff\ and \logg\ were $31000\pm$500\,K and 7.85$\pm$0.05, respectively, and can be seen in Figure~\ref{fig:J1519_spec_rvHalpha}. The black line is the single SDSS spectrum with the highest S/N shifted to the rest-frame, and the blue line is the best fit WD model spectrum. We did not use H$\alpha$ for the WD fit as the dC component contributes most to the spectrum in emission. In addition, we did not use the H9 line, as only half of the line is visible in the SDSS spectrum.

The WD temperature of our fit is in good agreement with that of \cite{Bedard2021}.  However, our \logg\ is 0.12 dex lower, resulting in both our WD mass and cooling age being lower than those  in \citet{Bedard2021}. For the purposes of this paper, we adopt our fit values of \logg\ and \teff. The WD properties we use can be found in Table~\ref{tab:wd_prop}, with the mass, radius, and cooling age coming from the models of \citet{Fontaine2001}.

\begin{deluxetable}{cccc}
\tablecaption{J1519 WD Properties}
\tablewidth{1.0\columnwidth}
\tablehead{\colhead{Parameter} & \colhead{Value} & \colhead{Error} & \colhead{Source}}
\decimals
\startdata
T$_{\rm{eff}}$ [K] & 31000 & 500 & (1)\\
$\log{g}$ [dex]& 7.85 & 0.05 & (1) \\
M [M$_{\sun}$]& 0.57 & 0.02 & (2) \\
R [R$_{\sun}$]& 0.015 & 0.001 & (2) \\
T$_{\rm{cool}}$ [Myr] & 7.7 & 0.2 & (2) \\
\enddata
\tablecomments{Best fit model parameters for the DA component of SDSS J151905.96+500702.9. Each parameter lists the source used: (1) this paper (2) from evolutionary models of \citet{Fontaine2001}} 
\end{deluxetable}
\label{tab:wd_prop}

\subsubsection{J1519 Radial Velocities}\label{subsec:J1519_mmtrvs}

Although RV variability has been detected in J1519, there are no published RV orbital fits for this system. Based on our photometric analysis, we found a period of $0.302356\pm0.000021$\,d ($\sim 7$\,hr) for J1519. Therefore, we conducted a spectroscopic monitoring of J1519 using the MMT spectroscopic setup as  was described in Section~\ref{subsec:spec_setup}. On the nights of 2020 August 19 and 20, we observed a sequence of 21$\times$200\,s exposures, on the night of 2020 August 22 we observed 27$\times$200\,s exposures, and on the nights 2021 April 21 and 23 we observed 24$\times$230\,s exposures. The exposures on each night were then co-added in threes, resulting in seven final epochs on the first two nights, nine epochs on the third night, and eight on each of the last two nights for a combined total of 39 epochs (with about 600\,s total exposure each), with an average S/N $\approx 5$ for all epochs in the continuum region near H$\alpha$. 

Since the full spectrum includes both stellar components, we measured the RV from the H$\alpha$ emission line, presumed to come from the dC atmosphere alone. First, for each epoch, we re-scaled the late-type (dCM) dC template of \citet{Roulston2020} to the flux in our MMT spectrum in the region of 6300--6500\,\AA. We then used this as the model for the dC continuum level of that epoch, which was used to calculate the H$\alpha$ emission line center, equivalent width and associated errors. The RV measurements have an average error of approximately 5~km~s$^{-1}$, and the equivalent width  measurements have an average error of approximately 0.18\,\AA.

\begin{figure*}
\centering
\epsscale{1.2}
\plotone{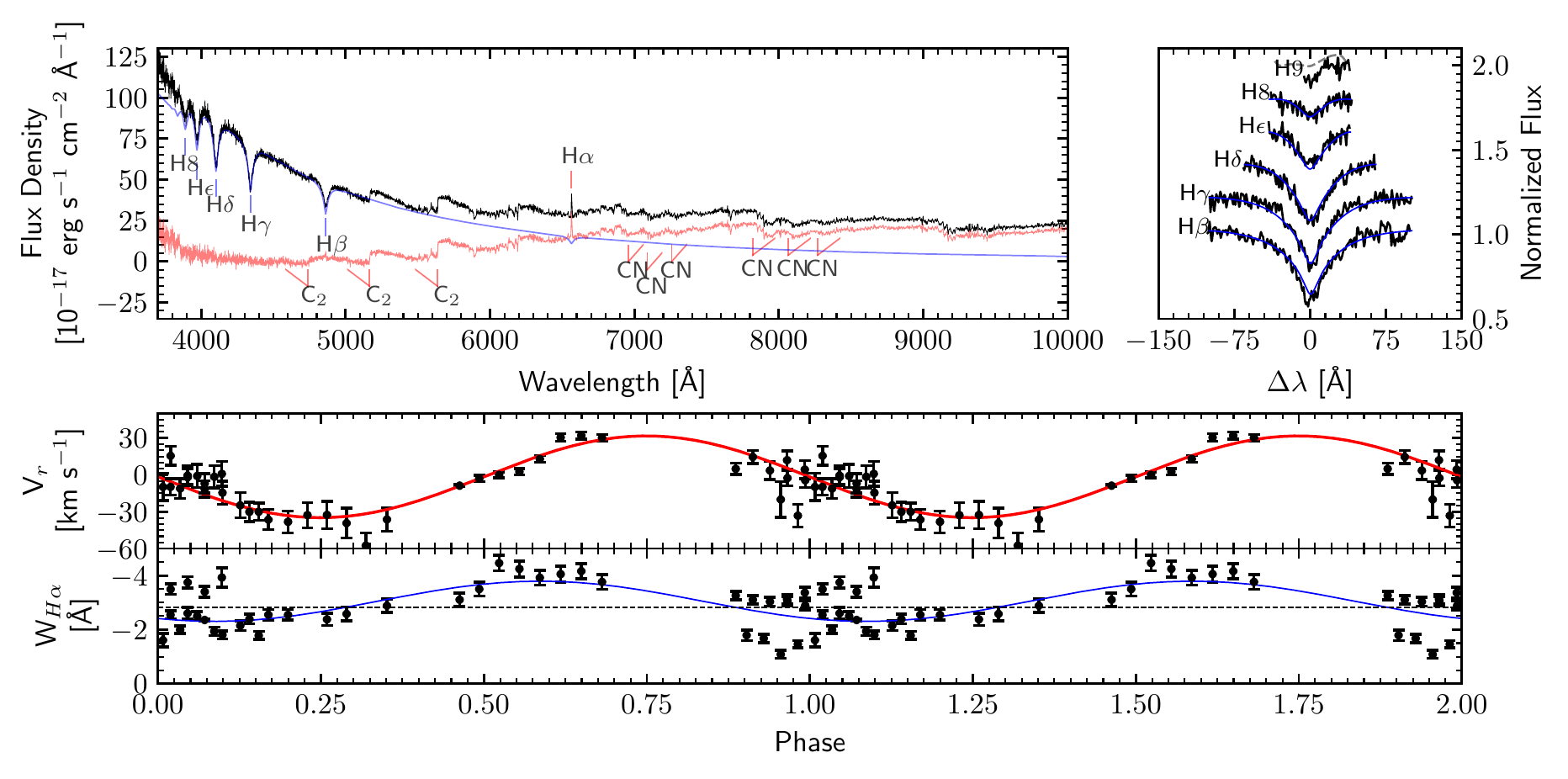}
\caption{ TOP: SDSS spectrum for the dC+DA SDSS J151905.96+500702.9, in black. The hot WD is visible, as are the carbon bands of C$_2$ and CN. The spectrum also shows H$\alpha$ emission. The best fitting model atmosphere to the DA component of the dC+WD composite J1519 is shown in blue, with the inset to the right showing a zoomed-in, stacked view of the Balmer lines used in the model fit (the black lines are the normalized flux of the WD component, and the blue lines are the best fitting model atmosphere, dashed grey lines means that that Balmer line was not used in the fit due to poor quality). The resulting dC component --- simply the observed spectrum with the WD model subtracted --- is shown in red. MIDDLE: RV as measured from the H$\alpha$ line for J1519, phased on the time of periastron passage from the RV fit, and the fit RV period of $0.327526$\,d. The red solid line represents the best fitting model. BOTTOM: Equivalent widths measured from the H$\alpha$ line, phased on the fit RV period of $0.327526$\,d. The blue curve is the best-fit model to the data of a single sinusoid. The y-axis has been inverted so that smaller equivalent width values (more emission) are up.}
\label{fig:J1519_spec_rvHalpha}
\end{figure*}

Figure~\ref{fig:J1519_spec_rvHalpha} shows the measured RV (middle) and H$\alpha$ equivalent widths (bottom) for J1519. To fit the RV curve, we used the \texttt{rvfit} program which uses a simulated adaptive annealing procedure, the details of which can be found in \citet{rvfit}. We left all parameters free to be fit, with the solution quickly converging to a circular orbit. We therefore refit the RV curve leaving all parameters free again except for the eccentricity, which we fix to $e=0.0$. The resulting best-fit model can be seen in Figure~\ref{fig:J1519_spec_rvHalpha} (red curve) and the fit parameters can be found in Table~\ref{tab:rvfits}.

The best fitting orbital period from the RVs ($0.327526\pm0.000012$\,d) is longer than the best photometric period by $0.025170\pm0.000024$\,d (about 36 minutes).  Fixing the period in the RV fitting procedure to that of the photometric period results in a poorer model fit, with the longer period model being a better fit at the 3.2$\sigma$ level. The best-fit semi-amplitude of $K_2 =33.3\pm1.4$~km~s$^{-1}$ ($\Delta RV_{\textrm{max}} = 2K = 66.6$~km~s$^{-1}$) is in agreement with the RV variations found by \citet{Whitehouse2018} and \citet{Roulston2019}, as their random epochs likely did not catch the true RV amplitude. However, the low measured semi-amplitude suggests an extremely low inclination of this system, with $i \approx 10$\arcdeg\ if we take our estimated dC mass of 0.41\,M$_\odot$ from Section~\ref{subsec:dc_bps_compare}. 

One possible explanation for a longer orbital period than photometric period is that J1519 was spun up by the accretion that it experienced, and has not yet synchronized the rotation and orbital periods in the approximate $8$\,Myr since mass transfer stopped (assuming the mass transfer ceased at the same time the WD formed). \citet{Green2019} analyzed \textit{Chandra} observations of J1519 (as well as five other H$\alpha$ emission dCs) and found it to show X-ray emission consistent with having a short rotation period, which would lend support to the accretion spin-up scenario. Deeper photometric imaging and RV follow-up, particularly of the WD component, could even better characterize this system. It is clear, however, that this dC has both photometric and RV variability on a $<$0.33-d timescale, indicating it most likely has a short orbital period and formed through a CE event.

\subsection{SDSS J123045.53+410943.8}\label{subsec:J1230}

\begin{figure*}
\centering
\epsscale{1.2}
\plotone{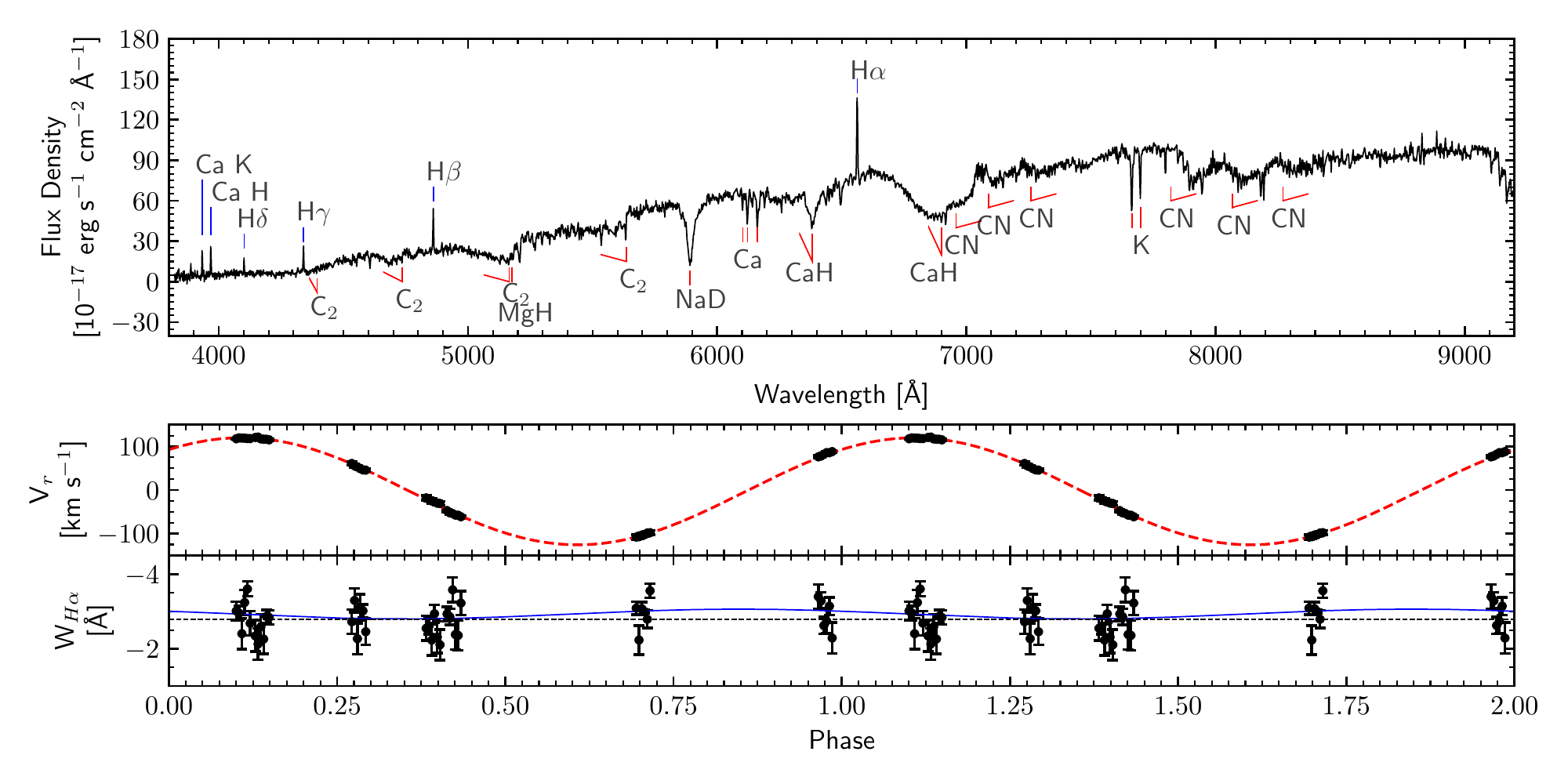}
\caption{TOP: SDSS spectrum for the dC SDSS J123045.53+410943.8. The typical carbon bands of C$_2$ and CN for late type dCs are labeled. Additionally, clear and strong emission of H$\alpha$, H$\beta$, H$\gamma$, H$\delta$, and Ca H and K are visible. Middle: RV as measured from the H$\alpha$ and $K\,I$ lines for SDSS J123045.53+410943.8, folded at the photometric period of $0.882519$\,d. The red dashed line represents the best fitting model. Bottom: Equivalent widths measured from the H$\alpha$ line, phased on the photometric period. The blue curve is the best fit single sinusoid model to the data, while the grey dotted line is the average equivalent width value. The y-axis has been inverted so that smaller equivalent width values (more emission) are up. We do not detect any significant variation in phase for the equivalent width, to a limit of $<$1.50\,\AA.}
\label{fig:J1230_spec_rvHalpha}
\end{figure*}

Another interesting dC is \object{SDSS J123045.53+410943.8} (J1230), whose SDSS spectrum is shown in Figure~\ref{fig:J1230_spec_rvHalpha}. J1230 shows the C$_2$ and CN lines typical of late type dC stars, but also shows strong absorption lines of K and a strong CaH band near 6800\,\AA. Additionally, this dC shows strong emission lines of H$\alpha$, H$\beta$, H$\gamma$, H$\delta$, and Ca H and K. Unlike J1519, there is no visible WD component in the spectrum of J1230. 

We observed J1230 ($r=16.6$\,mag) with the Binospec spectrograph on the 6.5m MMT using the setup described in Section \ref{subsec:spec_setup}. We took 6$\times$266\,s spectra on the nights of 2021 February 3, 5, 7, 8, 9, 11 and 15. This resulted in a total of 42 spectra, with an average S/N of 13 in the continuum region near H$\alpha$. We measured the line center and equivalent widths of the H$\alpha$ line, as well as for the two K lines visible in our MMT spectra of J1230. The emission and absorption lines have the same velocities, indicating the are coming from the same region. We average these velocities to measure the RV for each of the 42 epochs.

In the same method as J1519, we used the \texttt{rvfit} program to fit the RV curve of J1230. For this dC, we left all parameters free for fitting, with the resulting best period fit matching that of the photometric light curve ($P=0.882519\pm0.000020$\,d). We therefore fix the period to the photometric period and the eccentricity to 0.0, and refit the RVs. The resulting best fit can be found in Table~\ref{tab:rvfits} and the phased RV curve in Figure~\ref{fig:J1230_spec_rvHalpha}. As with J1519, we find the parameter errors using a MCMC method centered around the best fit parameters. 

The best fit gives a circular orbit with a semi-amplitude of $K_2 = 123.0\pm0.7$~km~s$^{-1}$. If we use our estimated dC mass from Section~\ref{subsec:dc_bps_compare} ($0.25$\,M$_{\odot}$) and an assumed WD mass of $0.6$\,M$_{\odot}$ the implied inclination of this system is around $i=56^\circ$.

The presence of multiple emission lines of H and Ca \deleted{may} suggest that the photometric variability of J1230 \replaced{may be}{is} coming from re-processing of the WD flux on the surface of the dC. Even though the WD companion to J1230 is not hot enough to be seen in the optical spectrum, it may be warm enough to still heat the surface of the dC. If this is true, we could expect the dC to be at maximum brightness when the WD-facing side is pointed toward us maximally, i.e. when the dC is moving transversely on the sky, between the ascending and descending nodes. Comparing the light curve of J1230 to the RVs however shows this is not the case, as if it were, we would expect the RV to be moving to the descending node after the photometric maximum, which the RV curve for J1230 is 0.33 out of phase with. This suggests that the  photometric variability \replaced{may}{is} not \deleted{be} coming from re-processing, but rather from spot rotation on an active dC (with the emission lines indicating chromospheric activity). We do note that the uncertainties on our epoch of maximum brightness and period \replaced{make it difficult for such precise comparison so many cycles after our light curve $t_0$, which could cause us to incorrectly predict the phase when our spectroscopy was collected in 2021~February by up to 0.75 cycles}{may cause us to incorrectly predict the phase when our spectroscopy was collected in 2021~February by up to 0.03 cycles}. \deleted{Follow-up photometry should improve the ephemeris and allow a more confident comparison.}

\subsection{SBSS 1310+561}\label{subsec:SBSS1310+561}

We observed \object{SBSS 1310+561} ($r=14.1$\,mag) using the FAST spectrograph on the 1.5m telescope at FLWO using the setup described in Section \ref{subsec:spec_setup}. We took 3$\times$300\,s spectra during the nights of 2021 February 10 and 11, and 6$\times$300\,s spectra on the night of 2021 February 12 for a total of 12 spectra with an average S/N of 32 in the continuum region near H$\alpha$.

Unlike with our MMT and Magellan observations, because of observing time constraints on our awarded FAST time, we chose to obtain these spectra close to the quadrature phases based on the ZTF photometry ($P=5.1878\pm0.0012$\,d). We assumed that the photometric period corresponds to the orbital period, and used t$_0$ from the light curve to calculate the expected times that  \object{SBSS 1310+561} should be at the quadrature phases ($\phi=0.25$ and $\phi=0.75$). Our actual observations were taken at phases $\phi=0.27\pm0.02$ and $\phi=0.47\pm0.01$.

From these spectra, we measured the RV at $\phi=0.27$ to be  $V_r =-79.7 \pm 9.5$~km~s$^{-1}$ and the RV at $\phi=0.47$ to be $V_r = -19.3 \pm 5.5$~km~s$^{-1}$. Taking the difference in these two velocities for this system ($\Delta RV = 60\pm11$~km~s$^{-1}$) can place a lower limit on the semi-amplitude of $K_{2} > 30 \pm 6$~km~s$^{-1}$. Using our estimated mass from Section~\ref{subsec:dc_bps_compare} ($0.46$M$_{\odot}$) and an assumed WD mass of $0.6$\,M$_{\odot}$, this constrains the inclination to $i \geq 25^\circ$ (if $i = 60^\circ$, then we would expect $K_2 = 109$~km~s$^{-1}$). Since the phase difference between our two epochs is quite small, this $\Delta RV$ suggests that SBSS 1310+561 is in a tight orbit and is very likely a PCEB.




\subsection{LAMOST J062558.33+023019.4}\label{subsec:J0625}
The dC \object{LAMOST J062558.33+023019.4} (hereafter J0625, $r=13.9$\,mag) was observed on the nights of 2021 January 11 and 12 using the Magellan MagE instrument setup described in Section \ref{subsec:spec_setup}. Each night, we observed 15$\times$300\,s exposures. The final reduced spectra consists of 30 epochs with an average S/N of 22 each in the continuum region near H$\alpha$. 

Using the H$\alpha$ emission line, we measured the RV of J0625 for each epoch. We found no evidence for RV variability, nor any variability in H$\alpha$ equivalent width. We found the RV to vary with only a standard deviation of 3.9~km~s$^{-1}$, and with a maximum $\Delta RV=12.1\pm3.2$ km s$^{-1}$. In addition, cross-correlation of the spectra across epochs resulted in no significant measured RV variations. Using our estimated mass from Section~\ref{subsec:dc_bps_compare} ($0.84$\,M$_{\odot}$) and an assumed WD mass of $0.6$\,M$_{\odot}$, this places a constraint on the inclination of $i \leq 7^\circ$ (if $i = 60^\circ$, then we would expect $K_2 = 44$~km~s$^{-1}$). This may suggest that the photometric variability is not related to the orbital period in this system since such a low inclination (and low semi-amplitude) is unlikely if the photometric period of $7.6080\pm0.0014$\,d represents the orbital period. Hence, this system adds weight to the evidence that the photometric variability in dCs may often be due to spot rotation.


 


\begin{deluxetable}{lcc}
\tabletypesize{\scriptsize}
\centerwidetable
\tablecaption{Radial Velocity Fits}
\tablehead{\colhead{Parameter} & \colhead{J1519} & \colhead{J1230}}
\startdata
$P$ [d]                   & $ 0.327526\pm0.000012$       & 0.882519$^\tablenotemark{a}$ \\
$T_p$ [MJD]               & $59080.2085 \pm 0.0053$       & 59265.07955 $\pm$ 0.00059 \\
$e$                       & $0.0^\tablenotemark{a}$    & $0.0^\tablenotemark{a}$ \\
$\omega$ [deg]            & $90.0^\tablenotemark{a}$   & $90.0^\tablenotemark{a}$\\
$\gamma$ [km s$^{-1}$]    & -1.7 $\pm$ 2.3              & -2.9 $\pm$ 0.5\\
$K_2$ [km s$^{-1}$]       & 33.3 $\pm$ 1.4               & 123.0 $\pm$ 0.7\\
 \hline 
$a_2\sin i$ [$R_\odot$]   & $0.2153\pm0.0090$          & 2.15 $\pm$ 0.01\\
$f(m_1,m_2)$ [$M_\odot$]  & $0.00125\pm0.00016$        & 0.170 $\pm$ 0.003\\
 \hline 
$\chi_\nu^2$              & 2.6                          & 0.97 \\
$N_{obs}$                 & 39                           & 42 \\
Time span [d]             & 247.2                         & 12.15\\
$rms_2$ [km s$^{-1}$]     & 11.6                          & 2.5
\enddata
\tablenotetext{a}{\footnotesize Parameter fixed during fitting.}
\tablecomments{Fit parameters from the radial velocity follow-up. The value for each parameter is given as the median of the marginalized distribution of the MCMC samples. The errors for each parameter are the 1$\sigma$ values from the marginalized distribution of the MCMC samples. Additionally, derived values for the orbital separation $(a\sin{i})$ and mass function $(f(m_1, m_2))$ are given. } 
\end{deluxetable}
\label{tab:rvfits}

\section{Common Envelope Connection}\label{sec:BPS}

For the progenitor of the dC companion to become a C giant, it must enter the third-dredge up phase \citep{Iben1974}. AGB stars have a degenerate CO core with a double-shell (moving outward from the core) of helium and hydrogen. As the hydrogen shell (which produces most of the energy) continues to fuse H into He, the helium shell surrounding the core continues to grow. Eventually, the helium shell experiences runaway fusion, driving expansion of the envelope material above. This He-shell ``flash'' and expansion means the star is now in the thermally pulsing AGB (TP-AGB) phase. Helium shell fusion causes the inter-shell region to become strongly convective, dredging helium fusion products to the surface, i.e., the third dredge-up. As the expansion continues, the pressure in the helium shell will drop, eventually stopping its energy production. The layers contract again with hydrogen shell fusion resuming, and the cycle repeats. 

Each successive thermal pulse becomes stronger, reaching deeper into the intershell zone, and the stellar radius increases \citep{Iben1983}. As helium shell fusion products are brought to the surface, it is possible that the envelope carbon abundance increases until C/O $> 1$. Since C preferentially binds with O, C$_2$ and CN bands only appear when C/O$> 1$, forming a C giant star. 

AGB stars going through the TP-AGB phase can reach radii of 800\,R$_\odot$ (3.7\,au) as they experience successively stronger thermal pulses \citep{Marigo2017}. Assuming an AGB mass of 2.5\,M$_\odot$, AGB radius of 800\,R$_\odot$, and a dC mass of 0.4\,M$_\odot$, this system would experience the beginning of a common-envelope (CE) with an initial period of $\approx 4.2$\,yr (if the dC mass is 1.0\,M$_\odot$ instead, then P$\approx 3.8$\,yr). Therefore, dCs with initial periods $\approx 4$\,yr (1500\,d) or less will very likely have experienced a CE phase, corresponding to the shorter-period peak modeled by \citet{Kool1995}. The dCs in this paper with P$< 1$\,d are most certainly the result of a CE spiral-in. Of the six dC periods in the current literature, two of them have P$< 3$\,d, so have likely experienced a CE. It seems then that many dCs may have experienced a CE phase.

\citet{DellAgli2021} recently studied the extreme AGB stars (those AGB stars which have extremely red mid-IR colors, e.g. \citealt{Gruendl2008}) and showed that \deleted{the} the excess dust and outflow densities of these stars may be explained by envelope stripping in a CE event. Their models suggest that these extreme AGB stars are actually post-common-envelope binaries (PCEBs) with orbital periods of order 1\,d, matching the periods for dCs in our sample. \citet{DellAgli2021} also found that the CE in their models starts after the rapid growth of the AGB radius, once the C/O ratio increases past unity, which corresponds well with the requirements for producing the short-period dCs we find. This makes these extreme AGB stars potential progenitors systems of the dCs that are in the CE phase currently. 

However, is mass accretion during a CE phase the most likely mass transfer mechanism to form dCs? We can address this question by looking at our periodic dC sample in the context of models that simulate expected binary populations.

\subsection{Binary Population Synthesis Models}\label{subsec:bps}
We used the binary population synthesis (BPS) models of \citet{Toonen2013} to see if the observed population of dCs can be reproduced by theory. The full details of the BPS models can be found in \citet{Toonen2013} and are briefly described here. 

These BPS models were created using the SeBa \citep{Portegies1996, Nelemans2001, Toonen2012, Toonen2013} population synthesis code. This code generates an initial population of binaries and simulates their evolution, taking into account processes such as stellar winds, magnetic breaking, mass transfer, common-envelope, and angular momentum loss. The initial stellar population is generated from the classical BPS distributions found in \citet{Toonen2013} via a Monte Carlo method. The resulting binaries are then convolved with a Galactic model including a star formation history that depends on time and location in the Milky Way based on \citet{Boissier1999} so that the simulated binaries can be compared to our observed sample.

For the synthetic populations used here, the common-envelope phase is modeled on the basis of the energy budget i.e. the classical $\alpha$-formalism of  \citet{Tutukov1979}. We discuss the results of two different models here that account for two different CE efficiencies: model $\alpha\alpha$ and $\alpha\alpha2$ which have $\alpha\lambda$ of 2 and 0.25, respectively. The parameter $\lambda$ is the structure parameter of the envelope to calculate the envelope binding energy \citep{Paczynski1976, Webbink1984, deKool1987, Livio1988, deKool1990, Xu2010}. 
The $\alpha$ parameter describes the efficiency with which orbital energy  is consumed to unbind the CE. A smaller value of $\alpha$ implies less efficient usage of orbital energy, and therefore a stronger shrinkage of the orbital period during the CE-phase. 
We do not consider the orbital angular momentum method of  \citet{Nelemans2000}, as this model does not reproduce the observed characteristics of the general PCEB (WD/main sequence) population  \citep{Toonen2013}.

Furthermore, the BPS models here allow for accretion during the common-envelope phase. The accretion rate is limited by the thermal timescale of the accretor times a factor that is dependent on the stellar radius and the corresponding Roche lobe \citep{Portegies1996, Toonen2012} following \citet{Kippenhahn1977, Neo1977,Packet1979,Pols1994}. The total accreted mass is then given by the integral of the accretion rate times the timescale of the CE event, which here is taken to be 100\,yr. This timescale is consistent with hydrodynamical simulations \citep{Ricker2008,Ivanova2013} and observations of hot subdwarf binaries \citep{Igoshev2020}, although cataclysmic variables may suggest a longer CE timescale, up to 10$^4$\,yr \citep{Michaely2019, Igoshev2020}. 



\subsection{BPS Comparison to Observed dC Sample}\label{subsec:dc_bps_compare}

We use the resulting model population for a direct comparison to our observed sample of short-period dCs, assuming the photometric period is the current orbital period. To do this, we estimated dC masses based on their infrared absolute magnitude M$_K$ in the $K$ band. Comparisons of M dwarf spectra \citep{Ivanov2004} to C star spectra  \citep{Tanaka2007} reveal them to be much more similar in the infrared than in the optical region. We used $K_s$ band (2.159\,\micron) magnitudes from the  Two Micron All-Sky Survey \citep[2MASS;][]{2MASS}. Six of our periodic dCs do not have $K_s$ band magnitudes. For these, we first fit {\em Gaia} absolute $G$ band ($M_G$) to the dCs that do have $K_s$ band magnitudes. This fit was then used to convert the {\em Gaia} M$_G$ into $M_{K_s}$ for the dCs lacking $K_s$ band magnitudes. We then fit M$_{K_s}$ for our dCs to stellar masses using data from \citet{Kraus2007}. This fitting also provides us with bolometric luminosities for the dCs in our sample. Comparing our bolometric luminosities to those provided in \citet{Green2019} (who used a spectral energy distribution method fitting $0.35-12.5$\,\micron) for the four dCs that overlap, we find our luminosities agree within 3\%, indicating our dC mass estimates should be reliable.

Our mass estimates can be found in Table~\ref{tab:LC_dc_gaia_prop}. We find that none of our dCs are fit with masses $> 1$\,M$_\odot$ or $< 0.2$\,M$_\odot$, in agreement with the range for which detectable C$_2$, CN, and CH bands are expected. We note that some of the lowest mass dCs may have been brown dwarfs or even planets before they accreted significant C-enriched material from their former AGB companion. 

Using the mass-radius relationship for main-sequence stars of \citet{Eker2018}, we estimate the radius for these periodic dCs as well, which are included in Table~\ref{tab:LC_dc_gaia_prop}. Using these estimated radii we calculate the Roche-lobe filling factor (RLFF), using the equation of \citet{Eggleton1983} to find the Roche radii. Six out of 34 of our periodic dCs may be experiencing RLOF back onto the WD (all have a RLFF $>1$ in Table~\ref{tab:LC_dc_gaia_prop}). However, we caution that physical parameters are derived from O-rich main-sequence models, which may not accurately represent all dCs. For example, (1) we do not know the mass of the unseen WD companion and assume it is $0.6$\,$M_\odot$ (2) we assume these mass-radius and M$_K$-mass relations hold for dCs, as they do for normal O-rich stars (3) dCs are thought to be of a lower metallicity population and studies have found that low metallicity M dwarfs may have smaller radii \citep{Kesseli2019} and (4) since dCs may have increased activity and magnetic fields due to their mass accretion, their radii may be inflated \citep{Kesseli2018}. We see no obvious evidence of flickering or accretion outbursts in any of our ZTF light curves that might indicate current RLOF back onto the WD.

\begin{deluxetable*}{ccDDDDDDDDhDhDD}
\centerwidetable
\tablecaption{Periodic dC Parallaxes, Distances, and Estimated Physical Parameters}
\tablehead{\colhead{R.A. (J2016.0)} & \colhead{Decl. (J2016.0)} & \twocolhead{$\varpi^\tablenotemark{{\tiny a}}$} & \twocolhead{$\sigma_{\varpi}^\tablenotemark{{\tiny a}}$} & \twocolhead{d$^\tablenotemark{{\tiny b}}$} & \twocolhead{$\sigma_{\textrm{d}}^\tablenotemark{{\tiny b}}$} & \twocolhead{BP - RP$^\tablenotemark{{\tiny a}}$} & \twocolhead{M$_{\textrm{G}}^\tablenotemark{{\tiny a}}$} & \twocolhead{M$_{\textrm{K}}$} & \twocolhead{M$_{\textrm{dC}}$} & \nocolhead{$\sigma_{\textrm{M}}$} & \twocolhead{$\log_{10}{(L_{\textrm{bol}} / L_\odot)}$} & \nocolhead{$\sigma_{\log{L}}$} & \twocolhead{R$_{\textrm{dC}}$} & \twocolhead{RLFF}\\ \colhead{ } & \colhead{ } & \twocolhead{$\mathrm{[mas]}$} & \twocolhead{$\mathrm{[mas]}$} & \twocolhead{$\mathrm{[pc]}$} & \twocolhead{$\mathrm{[pc]}$} & \twocolhead{$\mathrm{[mag]}$} & \twocolhead{$\mathrm{[mag]}$} & \twocolhead{$\mathrm{[mag]}$} & \twocolhead{[$\mathrm{M_{\odot}}$]} & \nocolhead{[$\mathrm{M_{\odot}}$]} &  \twocolhead{[erg s$^{-1}$] } &  \nocolhead{[erg s$^{-1}$] } & \twocolhead{[$\mathrm{R_{\odot}}$]}& \twocolhead{}}
\decimals
\startdata
00h47m06.76s & +00d07m48.80s & 0.68  & 0.18  & 1340 & 196 & 1.78 & 7.85  & 4.79                             & 0.67 & 0.00 & -0.88 & 0.00 & 0.59 & 0.06 \\
01h31m19.05s & +37d20m25.30s & 1.12  & 0.12  & 944  & 88  & 1.59 & 7.63  & 4.79                             & 0.67 & 0.00 & -0.88 & 0.00 & 0.59 & 0.27 \\
02h35m30.65s & +02d25m18.58s & 1.678 & 0.085 & 590  & 30  & 1.61 & 7.93  & 5.08                             & 0.60 & 0.00 & -1.05 & 0.00 & 0.52 & 0.22 \\
02h54m14.24s & +26d21m54.19s & 3.294 & 0.082 & 301  & 8   & 1.46 & 7.25  & 4.52                             & 0.74 & 0.00 & -0.72 & 0.00 & 0.67 & 1.10 \\
04h16m05.11s & +50d28m28.52s & 2.946 & 0.015 & 335  & 2   & 1.19 & 6.04  & 4.00                             & 0.87 & 0.00 & -0.40 & 0.00 & 0.83 & 0.12 \\
05h02m40.82s & +40d23m23.59s & 1.237 & 0.086 & 840  & 62  & 1.61 & 7.58  & 4.65                             & 0.71 & 0.00 & -0.79 & 0.00 & 0.63 & 0.13 \\
06h25m58.34s & +02d30m19.43s & 2.410 & 0.024 & 409  & 4   & 1.09 & 5.93  & 3.90                             & 0.90 & 0.00 & -0.34 & 0.00 & 0.86 & 0.11 \\
07h44m47.66s & +51d38m31.76s & 2.178 & 0.050 & 457  & 11  & 1.64 & 7.92  & 5.09                             & 0.60 & 0.00 & -1.05 & 0.00 & 0.52 & 0.23 \\
08h11m57.14s & +14d35m33.00s & 1.596 & 0.039 & 612  & 14  & 0.69 & 6.72  & 4.39                             & 0.77 & 0.00 & -0.64 & 0.00 & 0.70 & 0.45 \\
09h14m58.08s & +21d56m39.65s & 3.594 & 0.050 & 275  & 4   & 1.82 & 8.43  & 5.23                             & 0.56 & 0.00 & -1.14 & 0.00 & 0.49 & 0.25 \\
09h33m24.58s & -00d31m44.07s & 5.726 & 0.036 & 173  & 1   & 1.63 & 7.91  & 5.14                             & 0.59 & 0.00 & -1.08 & 0.00 & 0.51 & 0.27 \\
09h40m26.28s & +36d25m48.81s & 1.55  & 0.21  & 765  & 92  & 1.77 & 8.93  & 5.61                             & 0.48 & 0.00 & -1.35 & 0.00 & 0.41 & 0.17 \\
12h02m46.01s & +54d19m29.24s & 1.08  & 0.19  & 1103 & 170 & 1.98 & 8.87  & 5.32                             & 0.55 & 0.00 & -1.18 & 0.00 & 0.47 & 0.26 \\
12h08m53.35s & -00d08m47.99s & 0.78  & 0.37  & 2403 & 372 & 1.35 & 6.95  & 4.65$^\tablenotemark{{\tiny c}}$ & 0.71 & 0.00 & -0.80 & 0.00 & 0.63 & 0.70 \\
12h10m06.99s & +58d43m18.34s & 1.134 & 0.064 & 873  & 44  & 1.41 & 7.46  & 4.79                             & 0.67 & 0.00 & -0.88 & 0.00 & 0.59 & 1.04 \\
12h23m57.62s & +55d01m51.43s & 1.911 & 0.079 & 521  & 21  & 1.79 & 9.03  & 5.86                             & 0.43 & 0.00 & -1.48 & 0.00 & 0.36 & 0.50 \\
12h30m45.52s & +41d09m43.45s & 5.736 & 0.056 & 173  & 1   & 2.14 & 10.38 & 6.82                             & 0.25 & 0.00 & -1.96 & 0.00 & 0.22 & 0.20 \\
13h03m59.18s & +05d09m38.62s & 1.44  & 0.10  & 722  & 58  & 1.53 & 7.96  & 5.16                             & 0.58 & 0.00 & -1.10 & 0.00 & 0.50 & 0.20 \\
13h12m42.27s & +55d55m54.84s & 9.54  & 0.023 & 106  & 1   & 1.91 & 9.10  & 5.71                             & 0.46 & 0.00 & -1.40 & 0.00 & 0.39 & 0.08 \\
13h31m23.61s & +48d26m24.37s & 1.10  & 0.15  & 959  & 136 & 1.92 & 8.92  & 5.62                             & 0.48 & 0.00 & -1.35 & 0.00 & 0.40 & 0.75 \\
14h09m53.08s & -06d11m41.71s & 2.502 & 0.079 & 393  & 11  & 1.32 & 6.45  & 4.11                             & 0.84 & 0.00 & -0.47 & 0.00 & 0.79 & 0.87 \\
14h15m15.24s & +51d41m28.01s & 0.77  & 0.36  & 4420 & 799 & 1.19 & 6.62  & 4.47$^\tablenotemark{{\tiny c}}$ & 0.75 & 0.00 & -0.69 & 0.00 & 0.68 & 0.87 \\
15h11m44.58s & +38d59m10.46s & 2.05  & 0.11  & 487  & 29  & 1.84 & 8.90  & 5.88                             & 0.42 & 0.00 & -1.49 & 0.00 & 0.35 & 0.50 \\
15h15m42.72s & +52d01m45.47s & 1.291 & 0.065 & 775  & 37  & 1.61 & 8.00  & 5.30                             & 0.55 & 0.00 & -1.18 & 0.00 & 0.47 & 0.60 \\
15h19m05.93s & +50d07m03.14s & 2.274 & 0.064 & 437  & 14  & 0.77 & 9.08  & 5.95                             & 0.41 & 0.00 & -1.53 & 0.00 & 0.34 & 0.52 \\
15h24m34.12s & +44d49m55.84s & 0.96  & 0.19  & 1236 & 211 & 2.00 & 8.89  & 5.80$^\tablenotemark{{\tiny c}}$ & 0.44 & 0.00 & -1.45 & 0.00 & 0.37 & 0.62 \\
15h25m04.49s & +32d25m10.90s & 0.10  & 0.40  & 3280 & 598 & 1.55 & 7.42  & 4.91$^\tablenotemark{{\tiny c}}$ & 0.64 & 0.00 & -0.95 & 0.00 & 0.56 & 1.22 \\
15h30m59.26s & +45d12m00.33s & 2.215 & 0.044 & 444  & 8   & 1.81 & 8.45  & 5.39                             & 0.53 & 0.00 & -1.22 & 0.00 & 0.45 & 0.05 \\
15h35m32.92s & +01d10m16.22s & 1.18  & 0.39  & 1179 & 234 & 2.15 & 9.09  & 4.80                             & 0.67 & 0.00 & -0.88 & 0.00 & 0.59 & 1.07 \\
16h37m18.63s & +27d40m26.63s & 2.497 & 0.077 & 397  & 12  & 2.12 & 9.54  & 5.92                             & 0.41 & 0.00 & -1.51 & 0.00 & 0.35 & 0.21 \\
16h59m02.30s & +25d05m49.00s & 0.88  & 0.24  & 1293 & 220 & 2.07 & 8.87  & 5.79$^\tablenotemark{{\tiny c}}$ & 0.44 & 0.00 & -1.44 & 0.00 & 0.37 & 0.57 \\
19h23m55.93s & +44d58m32.20s & 1.238 & 0.048 & 802  & 28  & 1.36 & 6.90  & 4.79                             & 0.67 & 0.00 & -0.88 & 0.00 & 0.59 & 1.21 \\
22h08m10.01s & +25d17m30.17s & 4.121 & 0.053 & 241  & 3   & 1.56 & 7.75  & 4.79                             & 0.67 & 0.00 & -0.88 & 0.00 & 0.59 & 0.60 \\
23h41m30.74s & +15d19m43.20s & 0.27  & 0.12  & 2685 & 482 & 1.04 & 5.67  & 3.99                             & 0.87 & 0.00 & -0.40 & 0.00 & 0.83 & 1.60
\enddata
\tablenotetext{a}{\footnotesize From \citet{GaiaEDR3}}
\tablenotetext{b}{\footnotesize From \citet{GaiaEDR3_dist}}
\tablenotetext{c}{\footnotesize M$_{\textrm{K}}$ interpolated from M$_{\textrm{G}}$}
\tablecomments{Distances and magnitudes for the periodic dCs in our sample. We use the {\em Gaia} distances, colors, and magnitudes, as well as the 2MASS absolute $K$ magnitudes to estimate masses and bolometric luminosities for our dCs. For the solar bolometric luminosity, we adopt the value $\log_{10}{L_\odot}~=~33.58$. We also calculate the Roche-lobe filling factor (RLFF) under the assumption of a 0.6\,$\mathrm{M_{\odot}}$ WD companion. We calculate the mass errors to be of order 0.05\,$\mathrm{M_{\odot}}$, the $\log_{10}{(L_{\textrm{bol}} / L_\odot)}$ errors to be of  order 0.1, and the radius errors to be of order 0.05\,$\mathrm{R_{\odot}}$. However, we caution that physical parameters are derived from O-rich main-sequence models, which may not accurately represent all dCs.}
\end{deluxetable*}
\label{tab:LC_dc_gaia_prop}

To compare the BPS models directly to our observed dC sample, we applied a series of cuts and selection effects to the models, as follows: (1)  P $< 100$\,d (2) $r$ mag $< 19.5$ (3) M$_{dC} \le 1$\,M$_{\odot}$ (4) $1.0 < $M$_{\rm ZAMS} < 4$\,$M_{\odot}$ (5) the initial primary must be a TP-AGB star at the onset of the CE phase.  Here, M$_{dC}$ is the current mass of the main sequence companion in the BPS models, and M$_{\rm ZAMS}$ is the initial mass of the primary at the beginning of the models (which will become the AGB donor).

\begin{figure*}
\gridline{\fig{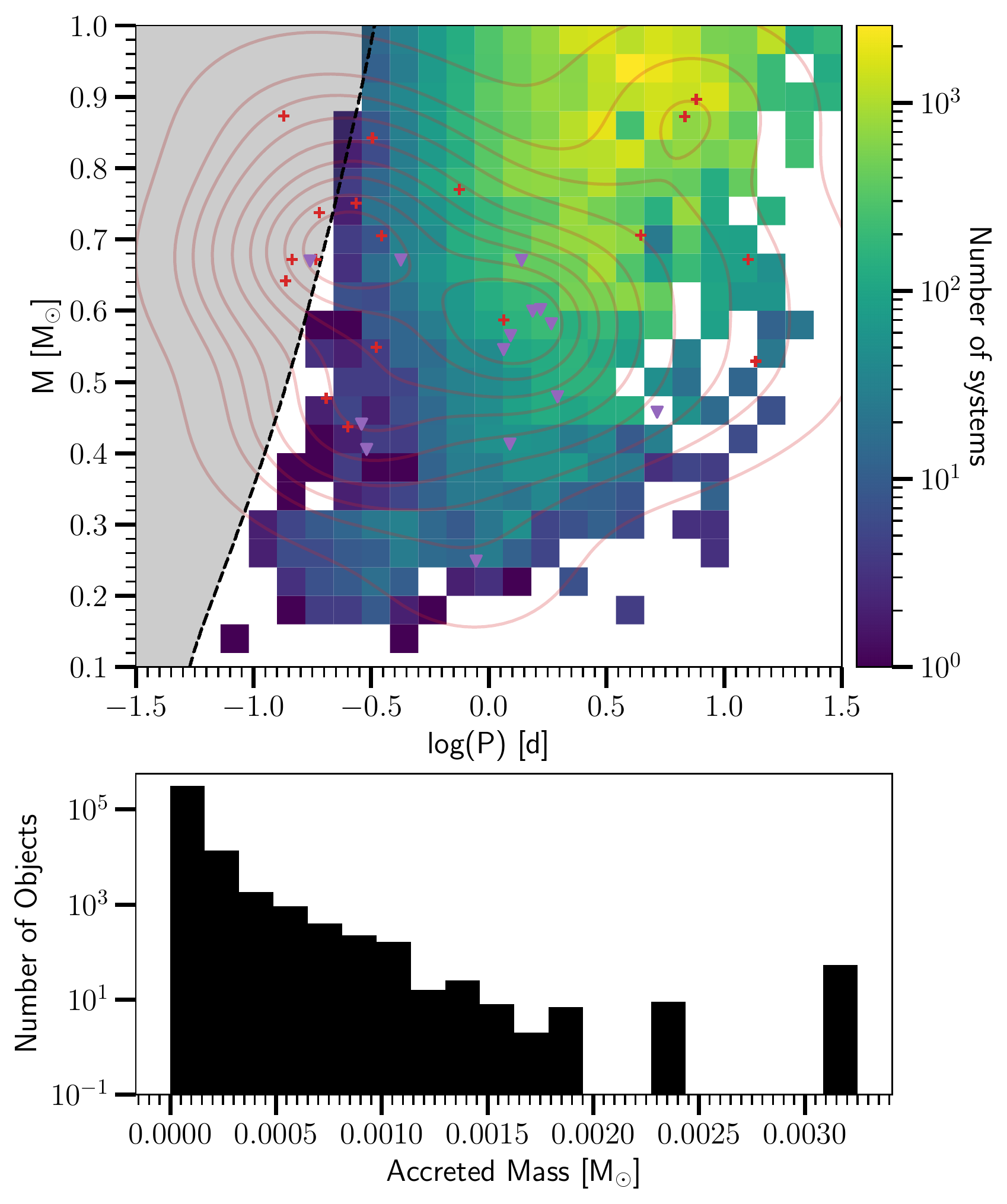}{0.35\textwidth}{(a)}
          \fig{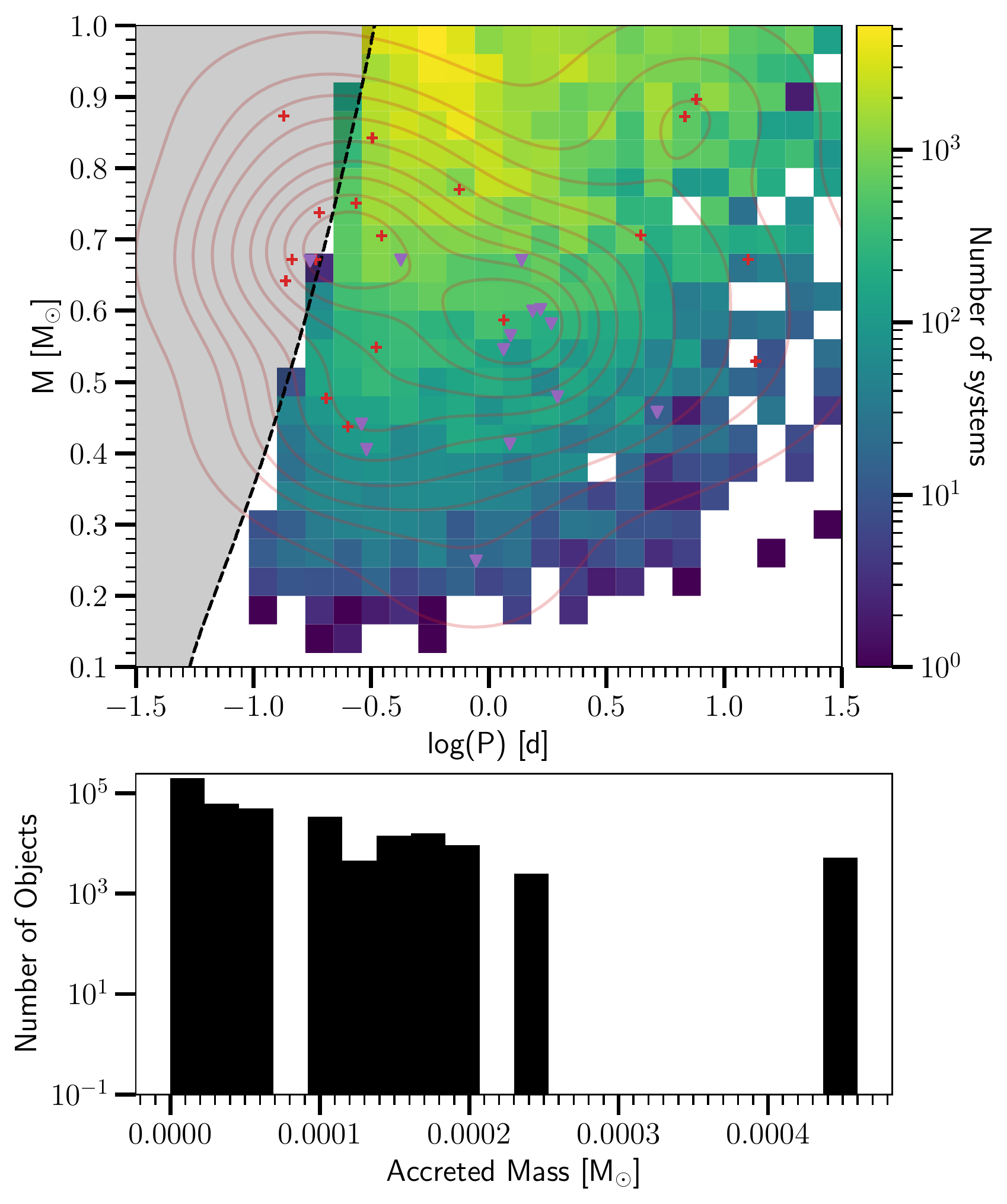}{0.35\textwidth}{(b)}
          \fig{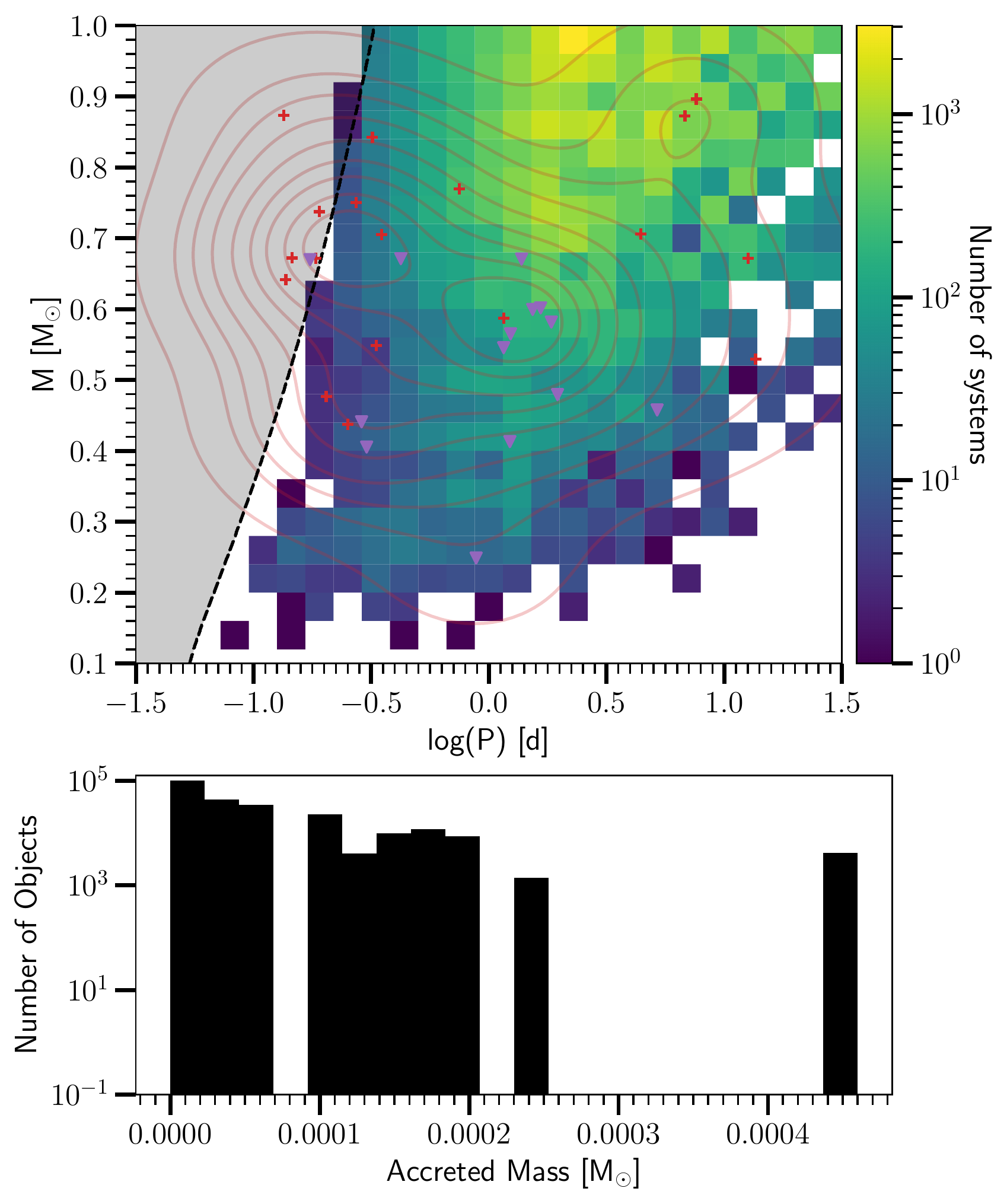}{0.35\textwidth}{(c)}}
\caption{Binary population syntheses results for model $\alpha\alpha2$, which is the model with lower common-envelope efficiency ($\alpha\lambda=0.25$) in which less orbital energy goes to unbinding the CE. The background heat map shows the number of systems (colored on a log scale) in model $\alpha\alpha2$ going through a CE phase while the primary is in the TP-AGB phase. The plotted masses are for the secondary (i.e. the dC) and periods are for the currently observed system. Each panel uses a different magnetic braking formalism: (a) magnetic braking from \citet{Rappaport1983} (b) magnetic braking from \citet{Ivanova2003} (c) magnetic braking from \citet{Knigge2011}. The red plus scatter markers are our observed short period dC sample without H$\alpha$ emission, while the purple triangles are dCs which show H$\alpha$ emission. The red contours are a KDE contour for the entire periodic dC sample. The dashed black curve represents the boundary for current dC systems to be experiencing RLOF, where systems to the left in the grey shaded region may be experiencing RLOF. The bottom panel shows a histogram of the estimated mass accreted by the secondary during the CE phase. While this model ($\alpha\alpha2$) reproduces the period distribution better than model $\alpha\alpha$ (see Figure~\ref{fig:model_aa}), the estimated accreted mass is even lower than that of $\alpha\alpha$ because of the lower CE efficiency. Since $>$0.03\,M$_\odot$ of mass transfer is likely required to create a dC, a CE is not likely to be the primary mechanism for accretion to form a dC.}
\label{fig:model_aa2}
\end{figure*}

\begin{figure}
\centering
\epsscale{1.2}
\plotone{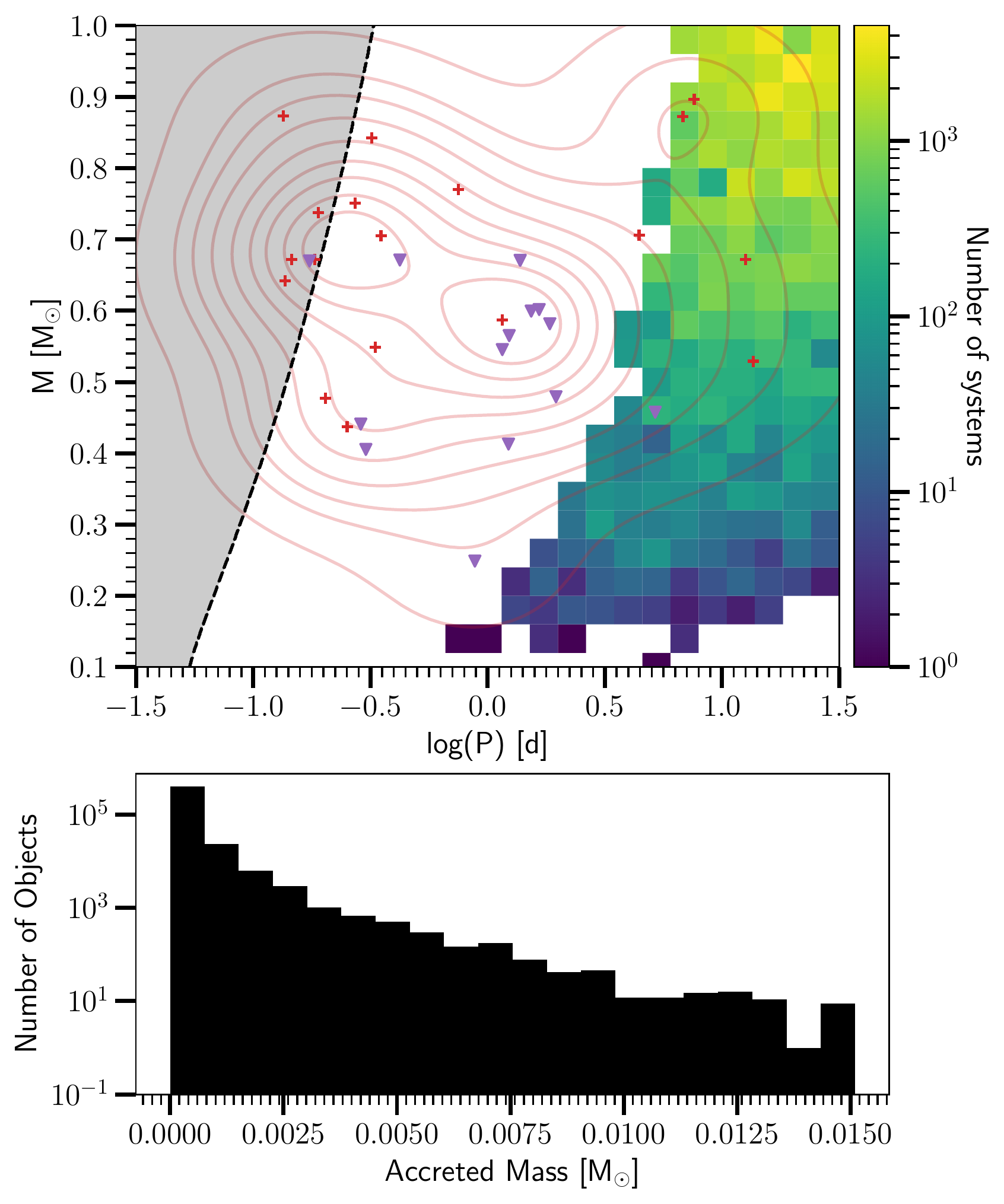}
\caption{Binary population synthesis results for model $\alpha\alpha$, which is the model with higher CE efficiency ($\alpha\lambda=2$). The background heat map shows the number of systems (colored on a log scale) in model $\alpha\alpha$ going through a common-envelope phase while the primary is on the TP-AGB phase. The plotted masses are for the secondary (i.e. the dC) and periods are for the currently observed system, including standard magnetic braking from \citet{Rappaport1983}. The red plus scatter markers are our observed short period dC sample without H$\alpha$ emission, while the purple triangles are dCs which show H$\alpha$ emission. The red contours are a KDE contour for the entire observed dC sample. The dashed black curve represents the boundary for current dC systems to be experiencing RLOF, where systems to the left in the grey shaded region may be experiencing RLOF. The bottom panel shows a histogram of the estimated mass accreted onto the secondary during the CE phase. This model ($\alpha\alpha$) does not reproduce our observed dC period distribution, as it does not produce periods below 1d, as well as not being able to accrete enough mass ($>$0.03\,M$_\odot$) we expect necessary to form dCs.}
\label{fig:model_aa}
\end{figure}

We show the resulting BPS models in Figure~\ref{fig:model_aa2} and Figure~\ref{fig:model_aa} ---  models $\alpha\alpha2$ and $\alpha\alpha$, respectively. In both figures, the BPS models are shown as the colored 2D histogram in mass and period (note that the histogram color scale is logarithmic and its range is different for each plot), and the periodic dC stars from this paper represented as the red scatter points (with KDE contours). The dashed black line represents the RLOF boundary, with systems occupying the region to the left (shaded in grey) filling their Roche lobes, under the assumption of a 0.6\,M$_\odot$ WD companion. Both figures also show a histogram of the estimated mass accreted during the CE phase (assumed to last 100\,yr).

Figure~\ref{fig:model_aa2} shows model $\alpha\alpha2$ ($\alpha\lambda=0.25$) and includes three different magnetic braking prescriptions. Panel (a) uses the magnetic braking of \citet{Rappaport1983}, panel (b) that of \citet{Ivanova2003} and panel (c) that of  \citet{Knigge2011}. Again, the color scale is logarithmic and its range is different for each sub-figure. 

Model $\alpha\alpha2$, however, does not reproduce the mass distribution of our dCs very well, generating low mass systems than observed (still under the assumption that our physical parameters derived from O-rich main-sequence models apply to dCs). While it may be that model $\alpha\alpha2$ does not produce low mass dCs, we have not considered our sample selection effects in this comparison. Our observed sample is likely biased toward lower mass dCs as (1) they have stronger C$_2$ and CN bands, and (2) their variability is fractionally larger and so easier to detect.

Model $\alpha\alpha$ (Figure~\ref{fig:model_aa}) uses a higher CE efficiency ($\alpha\lambda=2$) similar to classical BPS studies and includes the standard magnetic braking of \citet{Rappaport1983}. From Figure~\ref{fig:model_aa}, it is seen that this model is unable to reproduce the short  periods of our observed dC sample. This is in agreement with the conclusions based on the SDSS PCEBs (WD+MS systems; \citealt{Zorotovic2010,Toonen2013,Camacho2014}), where \citet{Toonen2013} found that standard efficiency  ($\alpha\lambda=2$) CE was also unable to reproduce the observed periods, as it generated too many long-period PCEBs.

A crucial shortfall is that the estimated mass accreted for all models is too small to convert a main-sequence star into a dC (see Section~\ref{sec:summary} for a discussion). 
\citet{Miszalski2013} suggest that to shift the secondary envelope from approximately solar (C/O)$_i \sim1/3$ to the observed (C/O)$_f >1$ requires $\Delta M_2 = 0.03$–$0.35\,M_\odot$ for $M_2 = 1.0$–$0.4\,M_\odot$. The predicted mass accretion in our BPS models is lower than this by 2$-$3 orders of magnitude.  Together with the strong mismatch between the modeled and observed dC period-mass distributions, it seems clear that there must be further mass accretion outside the brief CE phase.

Qualitatively it is also possible to argue that accretion during CE evolution is rarely significant for non-degenerate companions. The common envelope itself typically possesses much higher specific entropy than the surface of the accretor, with the consequence that matter accreted by the companion star reaches pressure equilibrium at the surface of that star with much higher temperature, and vastly lower density, than the accretors initial surface layer. A temperature inversion or roughly isothermal layer is expected to bridge this entropy jump with the result that, over the duration of the CE evolution (which is much shorter than the thermal time scale of the accretor), the accretor is thermally isolated from the common envelope, while the common envelope itself becomes increasingly tenuous.

A solution to the under-prediction of accreted mass may be that more mass accretion may take place before the CE phase, but during the TP-AGB phase, by wind accretion during wind-RLOF \citep[WRLOF;][]{Mohamed2007}. WRLOF is a mass transfer state that lies between standard wind mass transfer and standard RLOF, where the wind of the primary star is focused toward the secondary star. This results in increased mass transfer efficiency as compared to the standard Bondi-Hoyle-Lyttleton case \citep{Hoyle1939, Bondi1944}. 

In the WRLOF regime, the primary is technically not filling its Roche lobe. However, low-velocity wind matter is funneled through the Roche lobe to the companion, allowing for mass transfer to take place in binaries with wider orbits than the classical RLOF case. WRLOF would boost dC formation since, if the initial orbital separation is too small, the expanding AGB atmosphere can cause a CE before the third dredge-up can turn the AGB into a C star.

A variety of simulations \citep{Abate2013, Saladino2018, Saladino2019a, Saladino2019b} have shown that WRLOF in binaries with AGB primaries can have mass-transfer efficiencies of 40-50\%. For an average AGB wind mass loss rate of 10$^{-7}$--10$^{-4}$ M$_\odot$yr$^{-1}$ \citep{Hofner2015}, a main sequence companion could accrete enough material ($\sim 0.35$M$_\odot$) in only 10$^{3}$--10$^{6}$ yr, within the time an AGB star is expected to stay a C star \citep[10$^6$yr;][]{Marigo2017}. WRLOF has also been shown to efficiently tighten the orbit \citep{Saladino2018, Chen2018} so that more systems could be driven towards orbits with the periods we find.

Indeed, it appears the WRLOF may be the dominant mass transfer mechanism for many chemically peculiar stars. \citet{Abate2013} showed that simulations of AGB binaries with WRLOF were better able to reproduce the observed formation rates of the CEMP-s stars. \citet{Saladino2019a} and \citet{Saladino2019b} also performed simulations finding AGB binaries with WRLOF were consistent with the observed properties of the CEMP-s, CH, and Ba stars. This further strengthens the connection between dCs, WRLOF, and the other chemically peculiar stars.

However, detailed modeling of the WRLOF in the specific case of the TP-AGB phase with a C star is needed to understand how the larger stellar radii of AGB-C stars and the increased dust formation often found in their winds affect the WRLOF mass transfer efficiencies. The periodic dCs in this paper represent a prime sample that is ready for spectroscopic follow-up and for comparison to future models of WRLOF mass transfer.

\section{Summary}\label{sec:summary}

We searched ZTF light curves of a sample of 944 dCs for periodic signals. We found 34 dCs with signs of significant photometric variability, with 82\% having P$< 2$\,d. The most likely origins of this periodicity is either from spot rotation or surface heating of the dC from the close WD companion. Even if the detected ZTF periods arise from ellipsoidal variations and represent half the orbital period, such short periods are surprising for dC stars, which require significant accretion from a TP-AGB C giant to turn them into the C-enriched dwarfs we see today.

Spectroscopic follow-up is needed to determine the source of the variability in each of these periodic dCs, especially to confirm for the case of spot rotation that the system is circularized and tidally locked so that the rotation period can be assumed to equal the orbital period (i.e. that our reported photometric periods correspond to both rotational and orbital periods). The periodic dCs in this paper provide a rich new sample to target for spectroscopic follow-up, as well as to study dC formation and properties. We have confirmed the photometric period as the orbital period for one dC for which we have obtained spectroscopic follow-up. In two other dCs we have confirmed that they must have short orbital periods from their RVs (not confirming the photometric period as the orbital period however). In all three cases, these short (P$< 1$\,d) orbits indicate these dCs have indeed experienced a CE phase.

These short periods indicate that at least some dCs will experience a CE at some point in their formation, with P$< 1$\,d dCs having experienced experiencing substantial plunge-in. We used binary population synthesis models to show that the observed sample of dCs is not well-reconstructed by mass transfer during the common-envelope phase alone, since the dCs in our sample require at least 0.03\,M$_\odot$ of mass accretion but our models predict 2--3 orders of magnitude less transfer during the CE phase, suggesting mass accretion before the CE phase. However, some systems such as cataclysmic variables indicate CE timescales an order or two longer \citep{Michaely2019, Igoshev2020} than our assumed 100\,yr \citep[based on][]{Ricker2008, Igoshev2020}, which may substantially increase the amount of accreted material to the point that the CE alone could provide enough mass to form a dC. 

Hydrodynamical simulations of CE evolution typically find that accretion onto a non-degenerate companion is not common, because of the entropy barrier between the companion and the surrounding material \citep[e.g.][for a review]{Ivanova2013}. In fact, even in the case of neutron star companions, which can accrete more efficiently due to neutrino cooling, accretion is limited to $\lesssim0.1M_{\odot}$
\citep{MacLeod2015}. Further modeling of the CE phase involving C-AGBs may provide further insight.

dC systems that begin as very wide binaries would experience stable mass transfer and a widening of the orbit. Systems that initially are close would begin a CE phase either during the red giant branch or during the AGB before the TP-AGB and, without experiencing the third dredge-up during the TP-AGB, would not produce dCs. Therefore, it seems that the most likely mass transfer mechanism to form dCs is WRLOF. 

Further modeling of WRLOF in binaries with a TP-AGB star are needed to fully test this formation pathway of dCs. Additionally, further work is needed to understand the relationship between initial dC metallicity and mass to constrain the amount (and composition) of material that needs to be accreted to form a dC. This would be an important step in constraining the mass-transfer efficiency in the WRLOF case.


\facilities{FLWO:1.5m (FAST), Magellan:Baade (MagE), MMT (Binospec)}

\software{Astropy \citep{astropy1, astropy2},  Corner \citep{corner}, Matplotlib \citep{matplotlib}, Numpy \citep{numpy}, Scipy \citep{scipy}, Scikit-learn \citep{Scikit-learn}, TOPCAT \citep{topcat}}

\begin{acknowledgments} We gratefully acknowledge useful discussions with Jeremy Drake.
ST acknowledges support from the Netherlands Research Council NWO (VENI 639.041.645 grants).

Spectroscopic observations reported here were obtained at the MMT Observatory, a joint facility of the Smithsonian Institution and the University of Arizona.

Photometric light curves used herein are based on observations obtained with the Samuel Oschin 48-inch Telescope at the Palomar Observatory as part of the Zwicky Transient Facility project. ZTF is supported by the National Science Foundation under Grant No. AST-1440341 and a collaboration including Caltech, IPAC, the Weizmann Institute for Science, the Oskar Klein Center at Stockholm University, the University of Maryland, the University of Washington, Deutsches Elektronen-Synchrotron and Humboldt University, Los Alamos National Laboratories, the TANGO Consortium of Taiwan, the University of Wisconsin at Milwaukee, and Lawrence Berkeley National Laboratories. Operations are conducted by COO, IPAC, and UW. 

Funding for the Sloan Digital Sky Survey IV has been provided by the Alfred P. Sloan Foundation, the U.S. Department of Energy Office of Science, and the Participating Institutions. 

SDSS-IV acknowledges support and resources from the Center for High Performance Computing  at the University of Utah. The SDSS website is www.sdss.org.

SDSS-IV is managed by the Astrophysical Research Consortium for the Participating Institutions of the SDSS Collaboration including the Brazilian Participation Group, the Carnegie Institution for Science, Carnegie Mellon University, Center for Astrophysics $|$ Harvard \& Smithsonian, the Chilean Participation Group, the French Participation Group, Instituto de Astrof\'isica de Canarias, The Johns Hopkins University, Kavli Institute for the Physics and Mathematics of the Universe (IPMU) / University of Tokyo, the Korean Participation Group, Lawrence Berkeley National Laboratory, Leibniz Institut f\"ur Astrophysik Potsdam (AIP),  Max-Planck-Institut f\"ur Astronomie (MPIA Heidelberg), Max-Planck-Institut f\"ur Astrophysik (MPA Garching), Max-Planck-Institut f\"ur Extraterrestrische Physik (MPE), National Astronomical Observatories of China, New Mexico State University, New York University, University of Notre Dame, Observat\'ario Nacional / MCTI, The Ohio State University, Pennsylvania State University, Shanghai Astronomical Observatory, United Kingdom Participation Group, Universidad Nacional Aut\'onoma de M\'exico, University of Arizona, University of Colorado Boulder, University of Oxford, University of Portsmouth, University of Utah, University of Virginia, University of Washington, University of Wisconsin, Vanderbilt University, and Yale University.
\end{acknowledgments}

\bibliography{references}{}
\bibliographystyle{aasjournal}






\end{document}

%% file: LC_figset.tex
\figsetstart
\figsetnum{1}
\figsettitle{Periodic dC candidate light curves and power spectra}

\figsetend